\documentclass[]{acmart}
\usepackage{graphicx}
\usepackage{hyperref}
\usepackage{enumitem}
\usepackage{lipsum}
\usepackage{multirow}
\usepackage{tikz}
\usetikzlibrary{arrows.meta,positioning}
\usepackage{listings}
\usepackage{xcolor}
\usepackage[xcolor,rightbars]{changebar}

\usepackage{tcolorbox}
\tcbuselibrary{listingsutf8}
\usepackage{listings}
\tcbuselibrary{skins,breakable}
\usepackage{etoolbox}

\usepackage{booktabs}
\usepackage{tabularx}
\usepackage{array}
\usepackage{amsmath}

\newcommand{\refsec}[1]{Section~\ref{#1}}
\newcommand{\reftab}[1]{Table~\ref{#1}}
\newcommand{\refequ}[1]{Equation~\ref{#1}}
\newcommand{\reffig}[1]{Fig.~\ref{#1}}

\usepackage{pgffor}

\newcommand{\textitbf}[1]{\textit{\textbf{{#1}}}}

\newcommand*{\GitHubMain}[1]{\url{{https://github.com/onspatial/hd-gen}}}
\newcommand*{\GitHubProposedApproach}[1]{\url{https://github.com/onspatial/large-scale-dataset-generator#1}}

\tikzstyle{startstop} = [rectangle, rounded corners, minimum width=3cm, minimum height=1cm,text centered, draw=black, fill=red!30]
\tikzstyle{process} = [rectangle, minimum width=7cm, minimum height=1cm, text centered, draw=black, fill=blue!30]
\tikzstyle{arrow} = [thick,->,>=stealth]

\newtcblisting{mycode}{
  listing only,
  boxrule=0.3pt,
  arc=3pt,
  outer arc=3pt,
  boxsep=3pt,
  left=4pt,
  right=4pt,
  top=3pt,
  bottom=3pt,
after=\vspace{-0.3cm}
}

\renewcommand\footnotetextcopyrightpermission[1]{}

\begin{document}

\title{HD-GEN: A High-Performance Software System for Human Mobility Data Generation Based on Patterns of Life}
% \title{Large-Scale Human Mobility Data Generator, Calibrator, and Processor Based on Patterns of Life Simulation}

\author{Hossein Amiri}
\orcid{0000-0003-0926-7679}
\affiliation{%
    \institution{Emory University}
    \city{Atlanta}
    \country{USA}
}
\email{hossein.amiri@emory.edu}

\author{Joon-Seok Kim}
\orcid{0000-0001-9963-6698}
\affiliation{%
  \institution{Emory University}
    \city{Atlanta}
    \country{USA}
}
\email{joonseok.kim@emory.edu}

\author{Hamdi Kavak}
\orcid{0000-0003-4307-2381}
\affiliation{%
  \institution{George Mason University}
    \city{Fairfax}
    \country{USA}
}
\email{hkavak@gmu.edu}

\author{Andrew Crooks}
\orcid{0000-0002-5034-6654}
\affiliation{%
  \institution{University at Buffalo}
    \city{Buffalo}
    \country{USA}
}
\email{atcrooks@buffalo.edu}

\author{Dieter Pfoser}
\orcid{0000-0001-9197-0069}
\affiliation{%
  \institution{George Mason University}
    \city{Fairfax}
    \country{USA}
}
\email{dpfoser@gmu.edu}

\author{Carola Wenk}
\orcid{0000-0001-9275-5336}
\affiliation{%
  \institution{Tulane University}
    \city{New Orleans}
    \country{USA}
}
\email{cwenk@tulane.edu}

\author{Andreas Z{\"u}fle}
\orcid{0000-0001-7001-4123}
\affiliation{%
    \institution{Emory University}
    \city{Atlanta}
    \country{USA}
}
\email{azufle@emory.edu}

\renewcommand{\shortauthors}{Amiri, et al.}
\begin{abstract}
    Understanding individual-level human mobility is critical for a wide range of applications. Real-world trajectory datasets provide valuable insights into movement behaviors and patterns of life but are often constrained by data sparsity and participation bias. Synthetic data, by contrast, offers scalability and flexibility but frequently lacks realism.
To address this gap, we introduce a comprehensive software pipeline for generating, calibrating, processing, and visualizing large-scale individual-level human mobility datasets that combine the realism of empirical data with the control and extensibility simulations.
Our system consists of four integrated components:
\textbf{(1)} a data generation engine that constructs geographically grounded simulations using OpenStreetMap data to produce diverse mobility logs;
\textbf{(2)} a genetic algorithm--based calibration module that fine-tunes simulation parameters to align with real-world mobility characteristics;
\textbf{(3)} a data processing suite that transforms raw simulation logs into structured formats suitable for downstream applications; and
\textbf{(4)} a visualization module that extracts and presents key mobility patterns and insights from the processed datasets for improved interpretability.
Evaluation of generated trajectory datasets for the Atlanta, Georgia, USA region show realistic behavior that, despite emerging from a simulation without any reference to real human individuals, exhibits realistic human behavior that closely matches aggregate metrics of real-world datasets. 
We also provide a sensitivity analysis to study what simulation parameters affect simulation runtime. Code and simulated datasets are shared to provide the broad research community with large-scale dataset that, albeit not real, exhibit realistic human behavior while being orders of magnitudes larger than any open real-world mobility dataset.

\end{abstract}

\keywords{Trajectory Generation, Patterns of Life, Simulation, Trajectory Data}
\maketitle              
\pagestyle{plain}

% I keep this section commented to finalize the paper and have estimation fo the text in the paper. 
% \section*{Requirements}
% \input{content/requirements}

\section{Introduction}

Human mobility datasets are fundamental for understanding human behavior~\cite{toch2019analyzing, zhu2024generic,gonzalez2008understanding}, mobility patterns~\cite{mokbel2018mobility,schneider2013unravelling}, and traffic dynamics~\cite{chen2022lane,islam2021spatiotemporal}. They enable a wide range of applications, including outlier and anomaly detection~\cite{zhang2024large, zhang2024transferable, amiri2024urban}, infectious disease modeling~\cite{kohn2023epipol}, and urban planning and infrastructure design~\cite{isaacman2012human}. Mobility datasets can be broadly categorized as \textit{real} or \textit{synthetic}. Real world data, such as those collected from cell phone traces, provide high fidelity but raise serious quality and privacy concerns, while synthetic data offer improved scalability and privacy preservation at the cost of reduced realism \cite{Zufle2024,kim2024humonet}. In practice, privacy regulations~\cite{xiao2015protecting, khoshgozaran2011location, krumm2009survey}, de-identification techniques, and voluntary data collection introduce sparsity and information loss, resulting in significant trade-offs between data utility, realism, and accessibility~\cite{zhu2024synmob, amich2025exploring}.

To mitigate these limitations, researchers are increasingly relying on simulated trajectory datasets which provide scalable, noise free mobility information without privacy risks. However, many of the existing trajectory simulators rely on random or simplistic destination selection~\cite{duntgen2009berlinmod, brinkhoff2002framework} or parametric trip distributions~\cite{levandoski2012lars, armenatzoglou2013general}. While such approaches are suitable for benchmarking storage systems or query processing, we would argue that they fail to capture the intentional and routine driven nature of real human mobility.
The Pattern of Life (POL) simulation framework~\cite{kim2020location, zufle2023urban, amiri2024patterns} addresses this gap by modeling human needs, activities, and daily routines. POL generates purposeful mobility behaviors such as commuting, dining, and social interactions, producing trajectories that better reflect real world movement semantics (e.g., going to and from work). POL has been used to generate large scale datasets with realistic mobility patterns, social interactions, and check-in behavior. Prior work includes the generation of fine grained mobility and social network datasets across multiple U.S. cities~\cite{amiri2023massive}, datasets with injected anomalies for mobility anomaly detection~\cite{amiri2024urban}, and a GeoLife calibrated dataset optimized via genetic algorithms to balance realism and scalability~\cite{amiri2024geolife+}.

In this systems paper, we substantially extend our SIGSPATIAL 2025 short systems paper~\cite{amiri2025hd}, which introduced the software used in our previous studies~\cite{amiri2024patterns, amiri2024geolife+, amiri2023massive}, into a complete and systematic POL-based human mobility data generation system. This work is a direct extension of the short paper~\cite{amiri2025hd}, rather than a compilation of the separate studies cited above. Compared with the short paper, the present work is substantially expanded and includes significant new technical content, system components, and evaluation results.

Specifically, we introduce a unified software platform, referred to as HD-GEN, i.e., \textbf{H}uman mobility \textbf{D}ata \textbf{GEN}eration based on POL, that integrates trajectory generation, calibration, processing, and visualization into a coherent end-to-end pipeline. HD-GEN is a synthetic mobility data generation platform and is not intended to serve as a digital twin of a specific real-world population, individuals, environment, or transportation system. Rather, the data generated by HD-GEN aims at being realistic but not real.
%By exhibiting realistic mobility patterns but without compromising privacy of any real human.
The platform expands the functionality of POL while emphasizing reproducibility, scalability, and usability for large scale experimental and operational settings.

Building on the short paper, this extended version makes the following additional contributions:

\begin{itemize}
\item improves \textbf{reproducibility and usability} by introducing checkpointing mechanisms, providing practical guidance and management scripts, and offering Python-based tools that simplify configuration and execution.

\item introduces a detailed \textbf{visualization toolkit} for real-time analytics and monitoring that provides interpretable insights into simulated mobility patterns and system behavior.

\item enhances \textbf{scalability} for large-scale parallel execution on distributed and multicore systems, including support for calibration and checkpoint-based execution strategies.

\item provides \textbf{run-time improvements} demonstrated over 2,000 simulation instances with varying configurations to compare HD-GEN with the existing Patterns of Life simulation baseline.

%\item introduces a \textbf{realism analysis} that compares simulated outputs with real-world data to evaluate how accurately HD-GEN reproduces observed mobility patterns

\item introduces a \textbf{data validation} that analyzes the realism of the simulated data showing that aggregate metrics such as visited POI category counts and flows, daily traffic patterns, and distribution of radius of gyration are highly similar to real-world datasets and theoretical results.

\item delivers an improved \textbf{open-source repository} (\GitHubMain{}) with clearer documentation, execution scripts, and integrated visualization tools.

\end{itemize}

The remainder of the paper is organized as follows. In \refsec{sec:background_information}, we review relevant background information and related prior work. Next , in \refsec{sec:software_details}, we provide a technical overview of POL and describe the software design and system-level details of HD-GEN. In \refsec{sec:generated_dataset}, we summarize the datasets produced using HD-GEN. Finally, in \refsec{sec:conclusions}, we conclude by summarizing the contributions of this work and discussing directions for future research.

\section{Background Information and Related Work}
\label{sec:background_information}
In this section, we provide background information and related work on human mobility data generation, including an overview of the POL simulation and a discussion of existing trajectory datasets. We first describe the POL framework, its behavioral foundations, and its suitability for generating large-scale, realistic mobility data. We then briefly review commonly used real and synthetic trajectory datasets and highlight their limitations, motivating the use of POL and the need for calibration against observed mobility patterns.

% \repeattext[1]{Pol details}

\subsection{The Patterns-of-Life Simulation (POL)}

The POL simulation, introduced in Urban Life~\cite{zufle2023urban}, is a city-scale, agent-based modeling framework designed to generate realistic human mobility trajectories and interaction patterns. The simulation instantiates autonomous agents within an urban environment constructed from OpenStreetMap data, including road networks, residential zones,  workplaces, and recreational venues. Agents perceive and interact with this environment and each other while independently making decisions that reflect the routines and choices of individuals in everyday life.

Agent behavior in POL is governed by a combination of physiological, financial, and social needs inspired by Maslow’s hierarchy of needs~\cite{maslow1943theory}, together with goal-oriented decision making based on the Theory of Planned Behavior~\cite{ajzen1991theory}. At each simulation time step, agents evaluate competing needs such as earning income, acquiring food, maintaining social relationships, and participating in leisure activities. These needs are balanced against practical constraints including time availability, spatial accessibility, and personal resources. The resulting decisions determine when and where agents travel, producing realistic daily schedules and mobility patterns.
As agents move between different types of locations, the simulation generates fine-grained spatiotemporal trajectories as well as implicit social graphs that emerge from repeated co-location and interaction.  This design makes POL well suited for controlled experimentation and downstream analysis in mobility modeling, social network analysis, and urban studies.

\begin{figure}
    \centering
    \includegraphics[width=\linewidth]{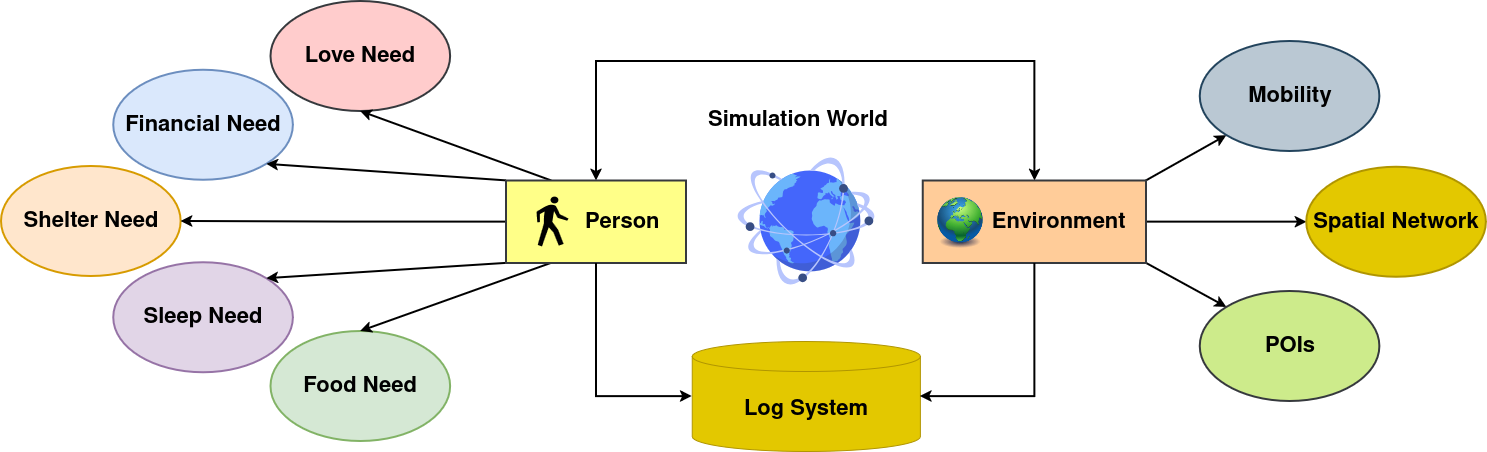}
    \caption{Interaction between a person agent and the simulation environment for generating human mobility logs.}
    \label{fig:pol-person-interaction}
    \Description{}
\end{figure}

\reffig{fig:pol-person-interaction} illustrates the interaction between a person agent and the simulation world. Each person is modeled as an autonomous agent with multiple needs, including food, sleep, shelter, financial safety, and love need from social connections. These needs evolve continuously over time and drive the agent’s decision making process. To satisfy its needs, the agent interacts with the environment, which defines mobility constraints, a spatial network, and points of interest such as workplaces, homes, and public venues. The environment determines where the agent can move and which actions are feasible at each location. Each agent perceives their environment, selects actions based on its current needs and context, and executes those actions through movement and interactions with points of interest, such as working at a workplace, or with other agents at recreational sites. All agent states, actions, and environment responses are recorded by the log system. The generated logs capture realistic spatiotemporal behavior and serve as the output of the simulation for downstream analysis and modeling.

\begin{figure}[t]
    \centering
    \includegraphics[width=1\linewidth]{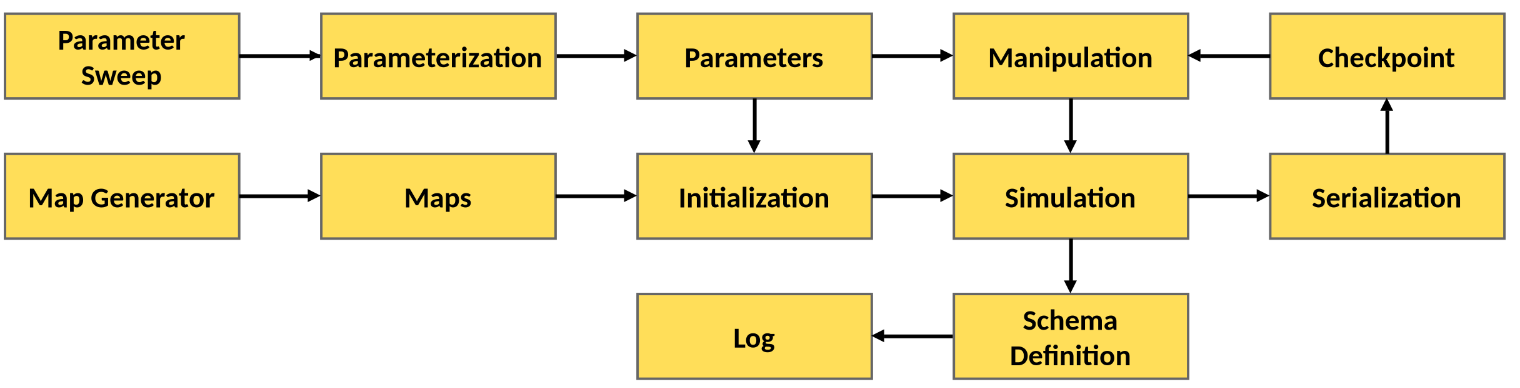}
    \caption{End to end POL simulation pipeline showing parameter configuration, map generation, initialization, execution, checkpointing, and structured log generation}\vspace{-0.4cm}
    \label{fig:pol-workflow}
\end{figure}

The POL simulation is implemented in Java utilizing the MASON toolkit~\cite{LukeMASONSimulationToolkit2018} and is configurable through a command-line interface, and publicly available on GitHub~\cite{zufle2023urban}. Users can specify city regions, population sizes, and behavioral parameters, enabling reproducible and extensible experiments. 
\reffig{fig:pol-workflow} presents the end to end simulation workflow. The process begins with a parameter sweep, where multiple configurations are defined to explore variations in agent behavior and environmental conditions. These configurations are formalized during parameterization, producing a consistent set of parameters that govern the simulation. In parallel, the map generator produces spatial representations of the environment, including networks and points of interest. The generated maps, together with the selected parameters, are passed to the initialization stage, where agents, environments, and initial states are instantiated. The simulation stage executes agent interactions over time based on the initialized state and active parameters. During execution, manipulation modules can dynamically modify parameters or system states, enabling interventions, scenario changes, or policy testing. The simulation state can be serialized to disk to implement checkpointing, allowing execution to be paused, restored, and reproduced in long running experiments. All simulation outputs are stored in structured formats guided by a predefined schema definition, which specifies how agent states, actions, and environmental events are recorded. The log system collects these records, producing standardized logs that serve as the final output of the simulation. 

In this system paper, we build directly on the original POL framework and preserve its behavioral and decision making logic while introducing new features and targeted improvements in execution efficiency and scalability. These optimizations enable simulations with larger populations and longer time horizons without altering the underlying semantics of agent behavior, ensuring comparability with prior work while substantially improving practical usability.

\subsection{Related Work and Relation to HD-GEN}

Human mobility data can be collected from real-world observations or synthesized using computational models. Real-world trajectory datasets capture observed mobility behavior and provide valuable insight into human movement and activity patterns. A widely used example is the GeoLife dataset~\cite{zheng2011geolife}, which contains four years of high-resolution GPS traces from approximately 180 users in Beijing, spanning both commuting and leisure activities. More recently, the large-scale YJMOB100K dataset~\cite{yabe2024yjmob100k} provides anonymized mobility records from 100,000 users at 30-minute intervals, aggregated into 500-meter spatial cells. While this scale enables population-level analysis, the spatial aggregation limits fine-grained, place-based behavioral inference. ~\cite{kashiyama2017open} further demonstrate how open government datasets and agent-based simulation can be combined to synthesize disaggregate, privacy-preserving people-flow trajectories for urban-scale mobility analysis.
Additional trajectory datasets focus on vehicles, including taxis~\cite{yuan2010t,PSG09} and buses~\cite{DC18}.  These datasets are valuable for modeling transportation systems but do not accurately represent individual human activity patterns, since each trip may involve different occupants and does not reflect continuous personal behavior. While real-world mobility data is often limited by privacy concerns, Open PFLOW provides a fully public, privacy-preserving synthetic people-flow dataset built from open data, enabling citywide mobility analysis with accuracy comparable to commercial datasets and traffic census data~\cite{kashiyama2017open}.
Synthetic trajectory datasets address limitations of real mobility data, including privacy constraints, proprietary access, and restricted scale. Recent deep generative approaches learn mobility patterns to produce realistic spatiotemporal trajectories, including factorized models such as EETG~\cite{zhang2022factorized} and diffusion-based methods~\cite{zhu2024synmob}. Optimization-based techniques further aim to enhance incomplete or sparse trajectory data~\cite{zhu2024generic,hu2022processing}. In contrast, the POL simulation framework~\cite{kim2020location,zufle2023urban,amiri2024patterns} generates trajectories driven by modeled human needs and decision processes rather than purely geometric or kinematic realism. 

In our prior work~\cite{amiri2023massive}, POL was used to generate a large-scale dataset exceeding 1.5 TB of mobility, check-in, and social interaction data. We further demonstrated the resulting data generator and its ability to reproduce key mobility and activity patterns in~\cite{amiri2024patterns}. Although the default parameter settings yield qualitatively realistic behavior, regional variation was not systematically characterized. This limitation motivated our subsequent study~\cite{amiri2024geolife+}, in which we calibrated POL to better match mobility patterns observed in the GeoLife dataset for Beijing. Finally, we integrated these efforts into a unified framework, HD-GEN, which was introduced in a concise system paper~\cite{amiri2025hd}.

\vspace{-0.2cm}
\section{Software Architecture and System Details}
\label{sec:software_details}
\vspace{-0.1cm}
This section describes the overall software architecture and execution flow of the proposed system. We first outline the software architecture, highlighting the main components and their interactions. We then detail the POL simulation, focusing on its technical design, scenario based testing with checkpointing, and real time log streaming. Finally, we present the HD-GEN pipeline, describing each phase of the workflow from data generation to calibration, processing, and visualization, and explaining how these stages integrate into a coherent end to end pipeline.

\vspace{-0.2cm}
\subsection{Software Architecture}
\label{sec:architecture}
\begin{figure}[t]
\centering
\includegraphics[width=\linewidth]{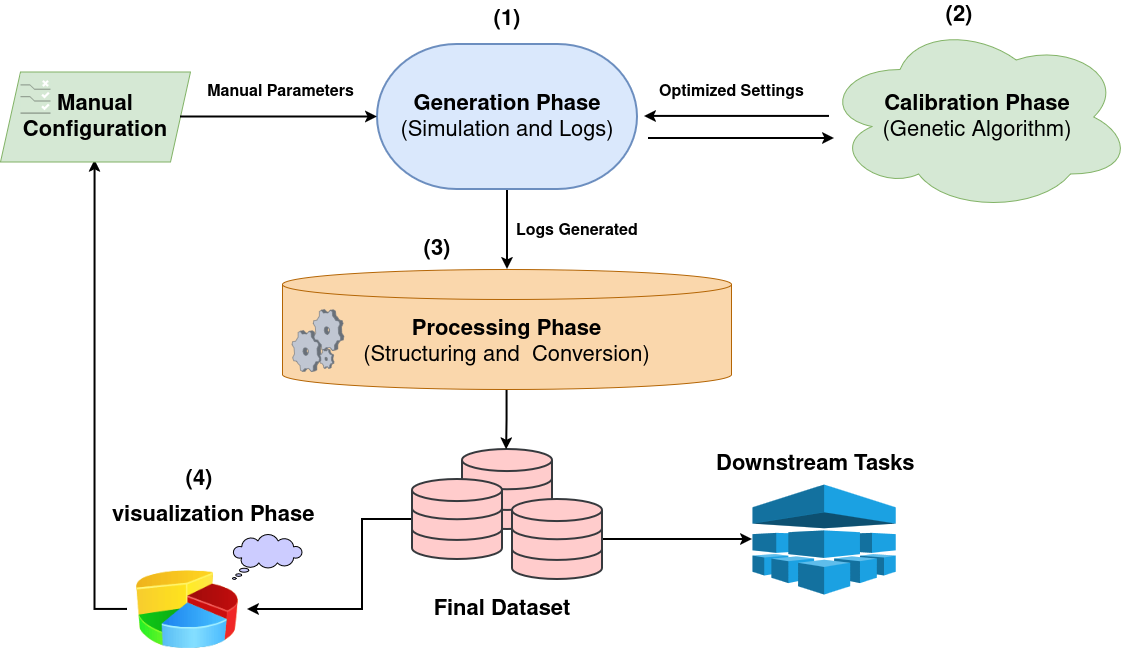}
\caption{Software architecture of HD-GEN.}
\label{fig:software-architecture}
\Description{}
\end{figure}

\vspace{-0.1cm}
Our system enables the generation of synthetic datasets that either statistically replicate real-world mobility patterns or conform to user-defined specifications. \reffig{fig:software-architecture} illustrates the architecture of the proposed software. The dataset generation pipeline is organized into four main phases: Generation, Calibration, Processing, and Visualization. The figure highlights the relationships among these phases and their interaction within the overall software framework. Each phase has its own internal sub-architecture, which is described in detail in~\refsec{sec:hd-gen-details}.

\textitbf{(1) Generation Phase}: The POL simulation is executed using a predefined set of parameters specified in an external configuration file. Each simulation instance runs on a single CPU core; however, the framework supports multi-core parallelism by executing multiple independent simulation instances concurrently. The simulator also provides checkpointing functionality, allowing execution to pause at designated checkpoints and resume from saved states with modified parameters. This enables efficient exploration of alternative hypotheses by branching multiple parallel simulation worlds from a common baseline state.

\textitbf{(2) Calibration Phase}: In this phase, key statistical metrics, implemented as scoring functions within the framework, are extracted from a reference real-world dataset and compared against the corresponding outputs of the simulation. A parallel genetic algorithm is implemented to optimize POL parameters by maximizing statistical similarity between simulated and observed data. Calibration is optional and may be skipped when fidelity to a reference dataset is not required or when users prefer to manually specify simulation parameters.

\textitbf{(3) Processing Phase}: The simulator generates multiple log files in fixed-size 512 MB chunks, including agent state, check-in, and social link records. These files are concatenated, and only the required columns are extracted and transformed into structured datasets suitable for downstream analysis. During this process, the simulation coordinate reference system, for example EPSG:26782, is automatically converted to EPSG:4326 to ensure compatibility with standard GIS tools. Users can also specify arbitrary input and output coordinate reference systems using the provided conversion tools, enabling transformation to EPSG:3857 or any other CRS required for metric-based analysis.

\textitbf{(4) Visualization Phase}: In this phase, the processed datasets are used for mobility analysis and visual analytics. This includes extracting key patterns such as trips, stay-point insights, check-in attributes, and movement clusters, as well as visual exploration of individual agents. These capabilities enable users to better understand mobility behavior, evaluate model performance, and gain insights into dataset characteristics.

\subsection{POL Details}

    This section describes the new features added to POL as part of this paper's extension, and the changes made to the original POL simulation to improve its performance.  
    Additionally, we explain the checkpointing, which allows for saving and resuming simulation states, and the log streaming feature that enables real-time monitoring of simulation progress.
    
    \subsubsection*{Performance Optimization}
    \label{sec:performance}

\begin{table}[t]
\centering

\caption{Performance comparison between Vanilla and Enhanced simulations for a 10 simulation day run. Initialization and simulation times are reported for different numbers of agents. A value of \texttt{-1} indicates that the simulation did not complete.
(\textbf{Init}: Initialization time; \textbf{Sim}: Simulation time; \textbf{Exit On/Off}: agent can/cannot exit; \textbf{s}: seconds; \textbf{m}: minutes; \textbf{h}: hours)}

\label{tab:performance-evaluation}

\begin{tabular}{|r|r|r|r|r|r|r|r|r|}
\hline
\multirow{2}{*}{Agents} &
\multicolumn{2}{|c|}{Vanilla Exit On} &
\multicolumn{2}{|c|}{Vanilla Exit Off} &
\multicolumn{2}{|c|}{Enhanced Exit On} &
\multicolumn{2}{|c|}{Enhanced Exit Off} \\

 & Init & Sim & Init & Sim & Init & Sim & Init & Sim \\
\hline
1000 & 33.2 s & 10.3 m & 30.7 s & 15.2 m & 8.7 s & 1.3 m & 4.9 s & 1.0 m \\ \hline
2000 & 1.9 m & 36.2 m & 1.4 m & 51.5 m & 11.6 s & 2.0 m & 11.1 s & 2.0 m \\ \hline
3000 & 2.6 m & 1.4 h & 2.2 m & 1.9 h & 19.1 s & 3.1 m & 20.9 s & 3.1 m \\ \hline
4000 & 4.2 m & 3.0 h & 4.1 m & 3.6 h & 35.3 s & 4.8 m & 35.6 s & 5.3 m \\ \hline
5000 & 4.6 m & 3.7 h & 5.7 m & 4.3 h & 48.8 s & 5.7 m & 53.0 s & 5.8 m \\ \hline
10000 & 13.9 m & 15.8 h & 13.0 m & 16.7 h & 3.3 m & 14.7 m & 3.3 m & 14.7 m \\ \hline
15000 & 24.1 m & 35.4 h & 22.8 m & 41.2 h & 8.7 m & 32.6 m & 8.0 m & 28.9 m \\ \hline
20000 &-1 &-1 &-1 &-1 & 12.8 m & 40.7 m & 11.2 m & 36.9 m \\ \hline
25000 &-1 &-1 &-1 &-1 & 18.9 m & 56.7 m & 19.3 m & 58.0 m \\ \hline
50000 &-1 &-1 &-1 &-1 & 1.1 h & 2.7 h & 1.0 h & 2.7 h \\ \hline
100000 &-1 &-1 &-1 &-1 & 4.4 h & 9.0 h & 4.5 h & 9.4 h \\ \hline
150000 &-1 &-1 &-1 &-1 & 9.6 h & 19.3 h & 9.8 h & 20.1 h \\

\hline
\end{tabular}
\end{table}

As we aim to simulate larger populations within the POL simulation, performance optimization becomes a critical concern. The original \emph{Vanilla} POL simulation exhibited limitations in scalability, particularly as the number of agents increased. To address these challenges, we implemented a series of performance optimizations in the \emph{Enhanced} POL simulation.
The enhanced simulation incorporates several key optimizations, including optimizing the initial parameter values, avoiding the logging of redundant or unnecessary information, and replacing the optimal travel distance algorithm with a Euclidean distance calculation~\cite{khan2024simplifying} to identify the nearest location.
%The original distance function, shown in \reffig{code:original-distance-function}, computes travel distance using an optimal pathfinding algorithm. While accurate, this method is computationally expensive, especially when called repeatedly by many agents. The modified version in \reffig{code:modified-distance-function} replaces this with a simple Euclidean distance calculation. This change greatly reduces computational cost, particularly in large-scale simulations where agents frequently evaluate distances to multiple locations.
These changes collectively contribute to a more efficient simulation process, reducing both initialization and execution times.

To evaluate the effectiveness of the proposed performance optimizations, we focus on the execution time of a single simulation instance. Since parallel execution does not alter the runtime of an individual simulation, this evaluation isolates the impact of the internal optimizations introduced in the enhanced system.
All experiments were conducted by simulating a period of 10 days (14{,}400 simulation ticks) while varying the number of agents. The experiments were executed on a machine equipped with an Intel Core i7-6700HQ CPU and 24~GB of RAM. We compare the baseline \emph{Vanilla} simulation against the \emph{Enhanced} simulation, which incorporates the proposed performance optimizations and we will use the \emph{Enhanced} simulation as the default simulation in HD-GEN going forward unless otherwise specified.

For each configuration, we report two distinct performance phases:
\begin{itemize}
    \item \textbf{Initialization time}, which measures the duration required to load spatial data, instantiate agents, initialize environments, and prepare simulation state.
    \item \textbf{Simulation time}, which measures the wall-clock time required to execute the full 10-day simulation period.
\end{itemize}

Results are reported for four execution modes: \emph{Vanilla} and \emph{Enhanced} simulations, each evaluated with agent exit conditions enabled (\textit{exit\_on}) and disabled (\textit{exit\_off}). The exit condition allows agents to leave the simulation when predefined criteria are met, reducing computational load in some scenarios. A value of \texttt{-1} indicates that the simulation did not complete within a reasonable time frame or was manually terminated due to excessive runtime.
\reftab{tab:performance-evaluation} summarizes the results across a wide range of agent counts. The results demonstrate that the \emph{Vanilla} simulation fails to scale beyond approximately 15{,}000 agents, frequently exceeding practical execution limits. In contrast, the \emph{Enhanced} simulation completes successfully for populations exceeding 150{,}000 agents, while substantially reducing both initialization and simulation times across all tested configurations. These results confirm that the proposed optimizations significantly improve scalability while preserving the behavioral semantics of the underlying simulation.

    \subsubsection*{Check-pointing}
    \label{sec:checkpointing}
    \begin{figure}
    \centering
    \includegraphics[width=0.9\linewidth]{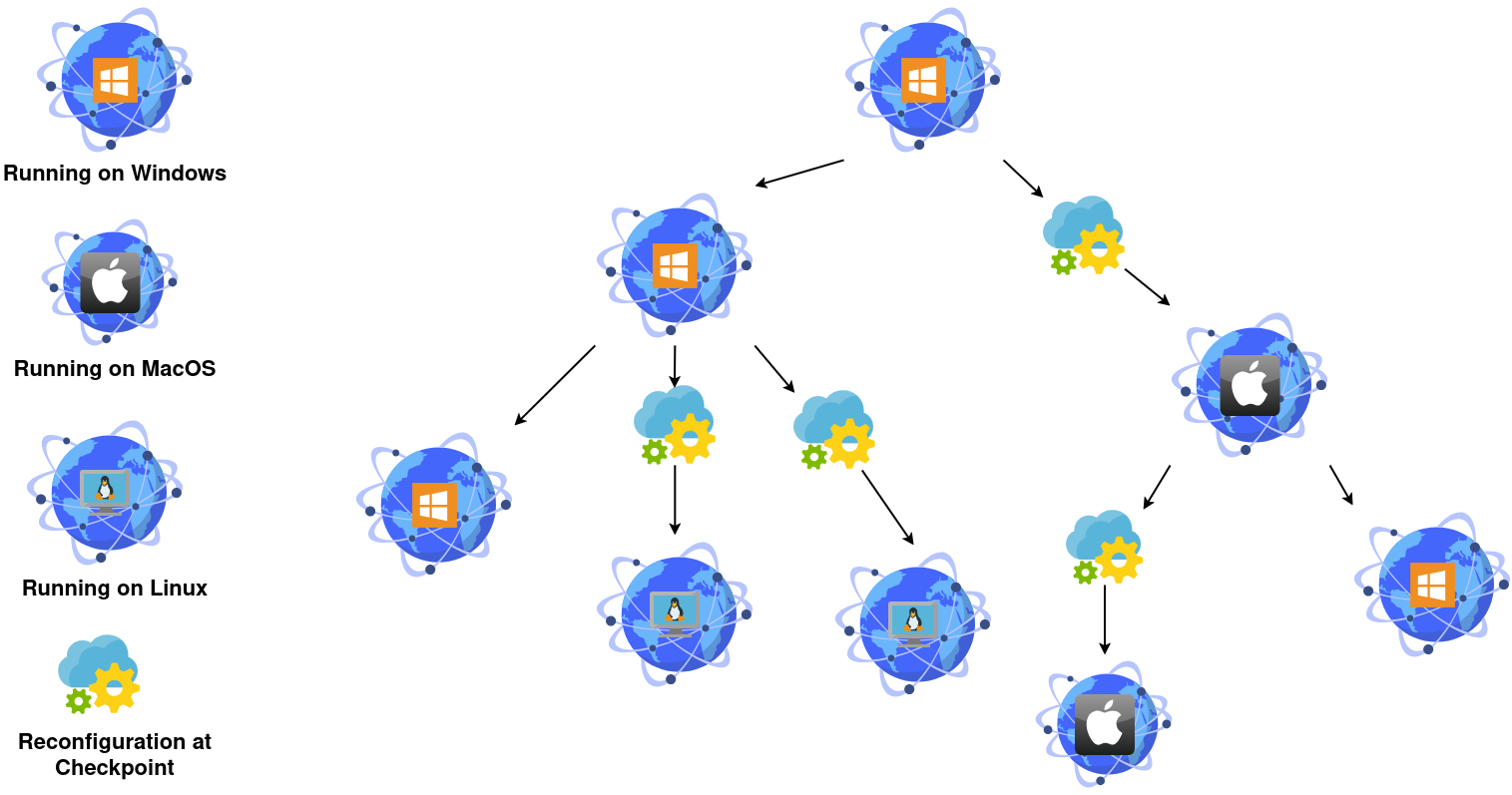}
    \caption{Deterministic checkpointing and cross platform execution of the simulation, showing how a paused run can be resumed and branched across Windows, macOS, and Linux machines to evaluate multiple scenarios from a shared state.}
    \label{fig:pol-checkpointing}
        \Description{}

\end{figure}

The POL simulation is deterministic, meaning that executing the simulation with the same input parameters always produces identical results. The simulation is also portable across different machines and operating systems, allowing execution to continue without breaking the pipeline or altering outcomes. These properties are essential for exploring multiple scenarios while preserving previously executed configurations.
Many studies require evaluating alternative scenarios under identical baseline conditions. For example, in an infectious disease setting, one may vary the number of initially infected individuals or compare different vaccination policies while keeping all other parameters fixed. To support such analyses, the simulation provides a checkpointing mechanism, illustrated in \reffig{fig:pol-checkpointing}. In this example, the simulation begins on a Windows machine and runs until a checkpoint is reached. At this point, execution is paused and the complete simulation state is serialized. From the same checkpoint, the simulation can be resumed with different configurations on multiple machines. For instance, one branch continues on a macOS machine with a modified configuration, while another resumes on Windows without reconfiguration. Additional branches are launched on Linux machines to evaluate further parameter variations. Each resumed execution represents a distinct scenario derived from the same initial state. All resulting runs generate logs that are stored in a unified format and remain fully compatible regardless of the execution platform. The underlying operating system is transparent to downstream processing, ensuring that all scenarios can be analyzed and compared consistently in subsequent analysis phases.

    \subsubsection*{Log Streaming}
    \label{sec:streaming}
    
\begin{figure}
    \centering
    \includegraphics[width=\linewidth]{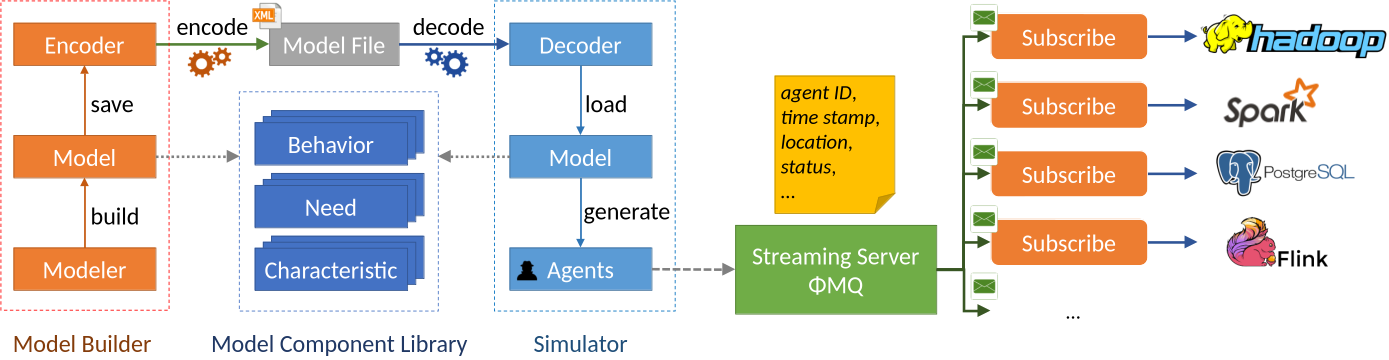}
    \caption{Real time streaming architecture for POL simulation logs, showing how agent events are published during execution and consumed by external analytics frameworks through a read only streaming interface.}
    \label{fig:pol-streaming}
    \Description{}
\end{figure}

Another aspect of the simulated world is real time log streaming, which enables interaction with simulation outputs while execution is in progress, rather than waiting for the simulation to complete and writing logs to static files. The POL simulation acts as the engine that generates agent behavior and produces event logs as scenarios evolve. A streaming server built on top of the logging system provides real time access to simulation data and publishes agent level events such as agent identifiers, timestamps, locations, and states. As shown in \reffig{fig:pol-streaming}, the simulator interacts with the POL model to satisfy agent needs and control behavior based on agent characteristics, while streaming events through a message broker to external systems. Downstream frameworks, including Hadoop, Spark, and PostgreSQL, can subscribe to these streams to consume simulation outputs in real time. The streaming interface follows a read only design and does not modify agent behavior or simulation state. Instead, it enables live observation, monitoring, and analysis of agent behavior as simulation time advances, supporting online analytics and early insight into emerging patterns.

\subsection{HD-GEN Details}
    \label{sec:hd-gen-details}

    \subsubsection*{ Generation Phase (1)}
    \label{sec:generation}

\begin{figure}
    \centering
    \includegraphics[width=0.9\linewidth]{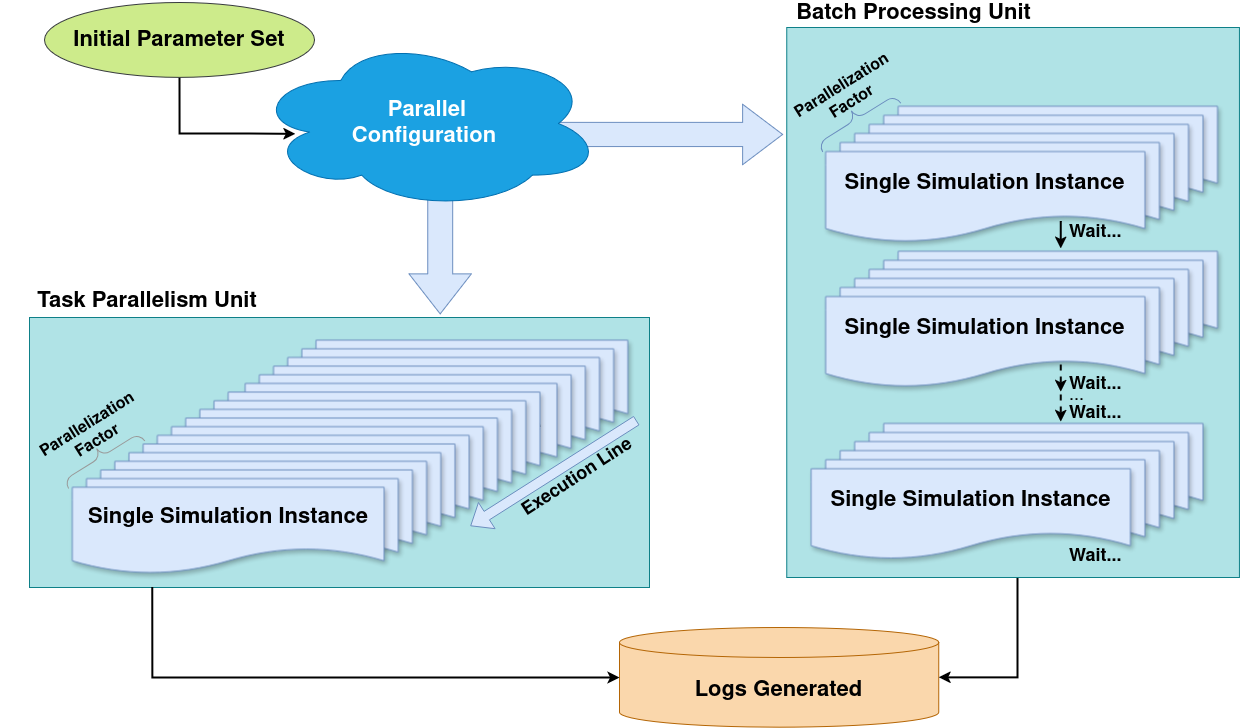}
\caption{HD-GEN simulation \textit{generation phase} architecture showing parallel configuration, task parallelism, and batch processing modes for scalable log generation.}
    \label{fig:hdgen-pipeline-generation}
    \Description{}
\end{figure}

To generate simulated data, we developed tools for regional map preparation, simulation execution, and large scale dataset generation in parallel. Map generation relies on OpenStreetMap (OSM) data accessed through the Overpass API \cite{overpass} to extract building footprints, road networks, and land use information. These raw features are further refined using predictive models to classify building usage. The map generation process produces three shapefiles, \textit{buildings.shp}, \textit{buildingUnits.shp}, and \textit{walkways.shp}, which collectively define the spatial environment used by the POL simulation. \reffig{fig:map-generator} shows an example script that generates a new map for the Minneapolis region. In addition to bounding box based queries, the tools support custom Overpass queries, allowing flexible selection of geographic regions and features for map generation.

\begin{figure} [h]
\centering 
\begin{mycode} 
bbx = [ -93.37271616619313, 44.87995237885596, -93.18917592484306, 45.065984365649186] 
output_folder = 'maps/minneapolis' 
pqgis.generate_map(bbx, output_folder, new_map=True) 
\end{mycode} 
\caption{Example script for generating a new simulation map using a bounding box and output path.} 
\label{fig:map-generator} 
    \Description{}
\end{figure}

\begin{figure}
    \centering
    \begin{mycode}
configured = get_configured_params("initial.json")
simulated = run(configured,batch_processing=False,parallel=8)
save_params(simulated,f"params.simulated.json")
    \end{mycode}
    \caption{An example of a Python script to run multiple simulation instances in parallel}
    \label{fig:run_parallel}
    \Description{}

\end{figure}
 
We also developed tools that allow users to define simulation parameters in a JSON file and run multiple instances in parallel, reusing calibration results when available. Two execution models are supported: batch-processing for dependent tasks (e.g., genetic algorithm tuning) and execution-line for independent runs (e.g., large-scale data generation). The number of parallel instances is configurable based on system capacity.
In \reffig{fig:run_parallel}, the Python script demonstrates how to execute multiple simulation instances concurrently. The function \texttt{get\_configured\_params} loads parameters from \texttt{initial.json}, allowing users to customize more than 60 default settings. Simulations are launched using \texttt{run()} with parallel execution enabled (\texttt{parallel=8}).
When \texttt{batch\_processing=False}, simulation instances are executed independently in an execution pipeline, which is suitable for large scale data generation. When \texttt{batch\_processing=True}, simulations are executed in groups, and execution waits until all instances in a group complete, which is useful for calibration and parameter tuning.
All results, including simulation outputs and execution metadata, are stored in \texttt{params.simulated.json} for subsequent analysis or reuse. This approach efficiently manages computational resources while preserving simulation integrity, enabling scalable data generation and systematic computational studies.

\reffig{fig:hdgen-pipeline-generation} illustrates the \textit{generation phase} architecture of HD-GEN. The process begins with an initial set of parameters, which are used to configure the simulation and create executable instances for each parameter set. Based on the value of the batch processing flag, simulation instances are dispatched to one of two execution modes. When batch processing is enabled, instances are sent to the batch processing unit, where simulations are executed in groups. Execution proceeds group by group, and each group waits until all simulations in that group complete before the next group is started. This mode is primarily used for calibration and coordinated experimentation. When batch processing is disabled, instances are sent to the task parallelism unit, where simulations are executed independently along an execution line. In this mode, execution is limited only by the parallel capacity of the system, making it suitable for large scale data generation. In both modes, all simulation runs produce logs that are collected and stored as the final output of the generation pipeline.
    
    \subsubsection*{Calibration Phase (2)}
    \label{sec:calibration}
    \begin{figure}
    \centering
    \includegraphics[width=0.8\linewidth]{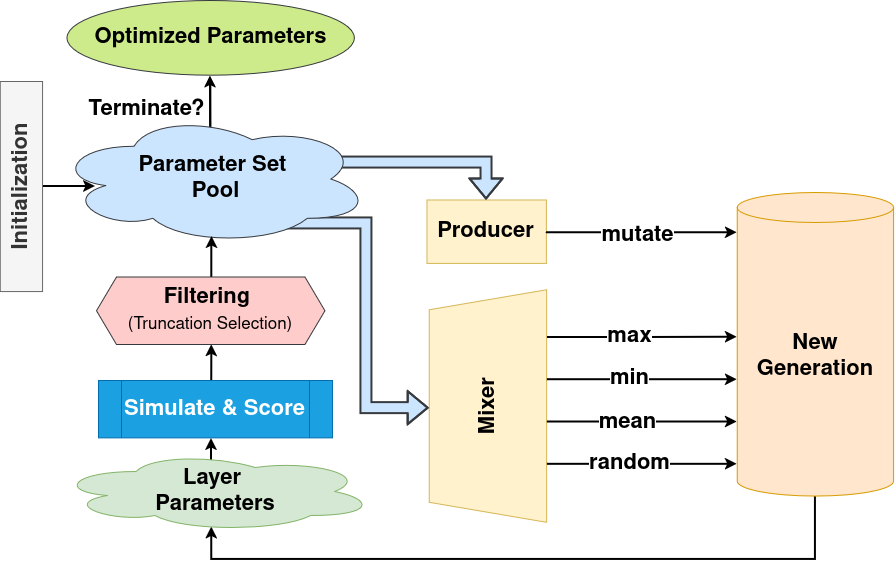}
\caption{Genetic algorithm workflow for obtaining optimized parameters. The initial parameter pool evolves through mutation (Producer) and mixing strategies (Mixer) to form new generations. After simulation and score evaluation, weak parameter sets are removed while strong ones are retained. The process continues until convergence, and the highest scoring parameters are selected}
    \label{fig:hdgen-pipeline-calibration}
        \Description{}

\end{figure}

For calibration, we used the GeoLife \cite{zheng2010geolife,zheng2011geolife} dataset as a real-world reference and applied a genetic algorithm guided by a similarity function to tune the simulation parameters. To keep the example simple, we compared real and simulated data using four metrics: average distance per trip (ADT), average distance per agent per day (ADA), maximum trip distance (MXD), and median trip distance (MDD). The similarity score in~\refequ{equ:score} quantifies the match, with values closer to 1 indicating greater similarity to the GeoLife dataset.

\begin{equation}
\label{equ:score}
\text{Similarity}(G,P) = 1 - \frac{1}{|M|} \sum_{k \in M} \frac{|k(P) - k(G)|}{k(G)}
\end{equation}

Here, $G$ is the GeoLife dataset, $P$ is the simulated dataset, $M = {ADT, ADA, MXD, MDD}$, and $k(\cdot)$ denotes the value of the metric.

To calibrate the simulations, we developed a genetic algorithm integrated into the data generation framework. We identified 63 parameters affecting agent behavior (e.g., number of interests, walking speed, rental cost ratio). A full list and documentation are provided in our GitHub repository (\GitHubMain{}). The goal was to find parameter values producing trajectories that best match GeoLife metrics.
The algorithm begins with simulations of $n$ using valid parameter values randomly chosen. Each run is scored using~\refequ{equ:score}, and the best performing parameter sets are retained as “parents.” New parameter sets (“children”) are then generated by combining parent values in five ways: maximum, minimum, mean, random mix, or random mutation. This process repeats until $n$ children are produced, each followed by a simulation and scoring step. New generations are iteratively created until convergence or manual termination. The best performing parameters across all generations are selected as the calibration result. The implementation is publicly available on GitHub.

\reffig{fig:hdgen-pipeline-calibration} illustrates the genetic algorithm process used to obtain optimized parameters. We start with an initial set of parameters that are placed into a pool. Two methods are then applied to generate new parameters:

\begin{enumerate}
    \item \textbf{Producer}: Creates new parameters by applying mutations. It uses the valid range defined for each parameter. For example, if the parameter joviality must be between 0 and 1, the producer will only generate values within that range.
    \item \textbf{Mixer}: Combines existing parameters using different strategies:
    \begin{itemize}
        \item \textit{Max}: selects the maximum value among the inputs. For example, if the values for a parameter $x$ are $1, 2, 3,$ and $4$, the mixed value becomes $4$.
        \item \textit{Min}, \textit{Mean}, and \textit{Random} apply their respective rules in a similar manner.
    \end{itemize}
\end{enumerate}

The results from the producer and mixer form the \textbf{new generation}. From this generation, a fixed number of parameter sets (equal to the layer size) are selected for the next stage.
After running simulations, each parameter set receives a score using the~\refequ{equ:score}. Any set with a score below a defined threshold is removed, and the surviving parameter sets are returned to the pool.
The algorithm continues iterating until a stopping condition is reached. Finally, the highest scoring parameter sets, such as the top one or top ten, are chosen as the \textbf{optimized parameters}.

\paragraph{\textbf{Calibration metrics}}

HD-Gen does not restrict calibration to a fixed set of mobility measures. Instead, the evaluation layer can be configured according to the requirements of the target application. In this system paper, we use four measures as a simplified example of the calibration procedure: ADT, ADA, MXD, and MDD. These measures provide an interpretable summary of typical travel distance, daily mobility intensity, extreme movement, and the central tendency of trip lengths. They are used only to demonstrate the calibration phase; the framework provides a broader collection of mobility, temporal, spatial, transition, semantic, and data-quality measures that can be selected or extended for a particular use case.
Let $\mathcal{T}$ denote the set of detected trips, $d_i$ the distance of trip $i$, and $\mathcal{Q}$ the set of active agent-day observations. For an agent-day observation $q\in\mathcal{Q}$, let $D_q$ denote the total distance traveled during that day. The four illustrative calibration measures are defined in~\reftab{tab:simplified-calibration-metrics}.

\begin{table}[t]
    \centering
    \caption{Simplified measures used to demonstrate the calibration procedure.}
    \label{tab:simplified-calibration-metrics}
    \small
    \begin{tabularx}{0.8\linewidth}{@{}l l X@{}}
        \toprule
        \textbf{Measure} & \textbf{Definition} & \textbf{Interpretation} \\
        \midrule
        ADT & $\displaystyle \frac{1}{|\mathcal{T}|}\sum_{i\in\mathcal{T}} d_i$
            & Mean distance of all detected trips. \\
        ADA & $\displaystyle \frac{1}{|\mathcal{Q}|}\sum_{q\in\mathcal{Q}} D_q$
            & Mean total distance across active agent-day observations. \\
        MXD & $\displaystyle \max_{i\in\mathcal{T}} d_i$
            & Longest detected trip in the dataset. \\
        MDD & $\displaystyle \mathrm{median}_{i\in\mathcal{T}}(d_i)$
            & Median distance of all detected trips. \\
        \bottomrule
    \end{tabularx}
\end{table}

Before the measures are calculated, mobility records are standardized and ordered by agent and time. Calendar date, hour of day, day of week, and weekend status are derived from each timestamp. Consecutive observations are linked to obtain elapsed time and traveled distance. When a distance value is unavailable, it can be reconstructed from consecutive geographic coordinates using the great-circle distance. Invalid transitions, non-positive time intervals, distances below the minimum movement threshold, and implausibly large distances are excluded from trip-level calculations. Locations are represented using an available place identifier or, when such an identifier is absent, by assigning coordinates to a spatial grid. These steps provide a consistent basis for comparing real and simulated mobility traces.
In this extension, we expand the set of evaluation metrics, as summarized in~\reftab{tab:available-calibration-metrics}. For measures represented by a sample of observations, the framework reports the number of observations, mean, standard deviation, minimum, selected percentiles, median, and maximum. Consequently, calibration can combine scalar measures, such as mean trip distance, with distributional measures, such as the complete trip-distance distribution.

\begin{table*}[t]
    \centering
    \caption{Mobility measures currently available as a calibration blueprint in HD-Gen. The selected subset may be adapted to the requirements of a specific application.}
    \label{tab:available-calibration-metrics}
    \small
    \begin{tabularx}{\textwidth}{@{}>{\raggedright\arraybackslash}p{0.16\textwidth} X X@{}}
        \toprule
        \textbf{Category} & \textbf{Available measures} & \textbf{Calibration purpose} \\
        \midrule
        Dataset coverage
        & Number of agents, observations, detected trips, unique locations, covered days, and temporal coverage.
        & Verifies that the simulated dataset has a scale and observation period comparable to the reference dataset. \\

        Data quality
        & Number of distance outliers, share of stationary or zero-length transitions, non-positive temporal gaps, and missing distances.
        & Detects invalid or implausible records before similarity is evaluated. \\

        Distance and mobility volume
        & Total distance, average distance per trip, distance per observation, per-trip distance statistics, per-agent total distance, per-agent-day total distance, and total distance per day.
        & Captures overall mobility intensity, typical trip length, inter-agent variability, and day-to-day variation. \\

        Visitation activity
        & Observations, trips, and active days per agent; observations and trips per agent-day; locations visited per agent-day; unique locations per agent; and home-visit share.
        & Measures activity frequency, regularity, location diversity, and the relative importance of the inferred home location. \\

        Spatial exploration
        & Radius of gyration, ranked exploration distribution, location entropy, normalized location entropy, and the visit share of the most frequent one and three locations.
        & Quantifies the spatial extent of movement and the balance between routine behavior and exploration. \\

        Temporal behavior
        & Time between consecutive observations, trip-duration gap, estimated speed, active time span per agent-day, number of active agents per day, weekend share, night-time share, hourly distribution, and day-of-week distribution.
        & Evaluates daily and weekly rhythms, activity duration, temporal sparsity, and plausible movement timing. \\

        Movement transitions
        & Number of unique origin--destination pairs, transition entropy, share of the most frequent transition, and the most frequent origin--destination movements.
        & Assesses the diversity and concentration of repeated movement patterns. \\

        Empirical distributions
        & Distributions of locations visited per agent-day, exploration rank, trip distance, daily distance per agent, and radius of gyration.
        & Supports distribution-level calibration rather than relying only on averages or extrema. \\

        Semantic behavior
        & Number of place categories, place-category entropy, most frequent categories, category diversity per agent-day, and frequent transitions between place categories.
        & Evaluates activity semantics when place-category information is available. \\
        \bottomrule
    \end{tabularx}
\end{table*}

For a selected set of calibration measures $M$, the similarity score in~\refequ{equ:score} averages their relative absolute errors. A score of $1$ represents an exact match, while lower values indicate larger deviations from the reference dataset. Because the formulation is not bounded below, sufficiently large deviations may produce negative scores. In addition, every reference value used in the denominator must be positive; otherwise, a zero-safe normalization should be introduced.
The metric set is therefore application dependent. A transport-oriented study may emphasize trip-distance and origin--destination distributions, whereas an activity-based study may assign greater importance to temporal rhythms, location diversity, or semantic place categories. Additional user-defined measures can be incorporated into the same evaluation stage and included in the genetic-algorithm objective without changing the overall calibration workflow.
    
    \subsubsection*{ Processing Phase (3)}
    \label{sec:processing}
    
\begin{figure}
    \centering
    \includegraphics[width=\linewidth]{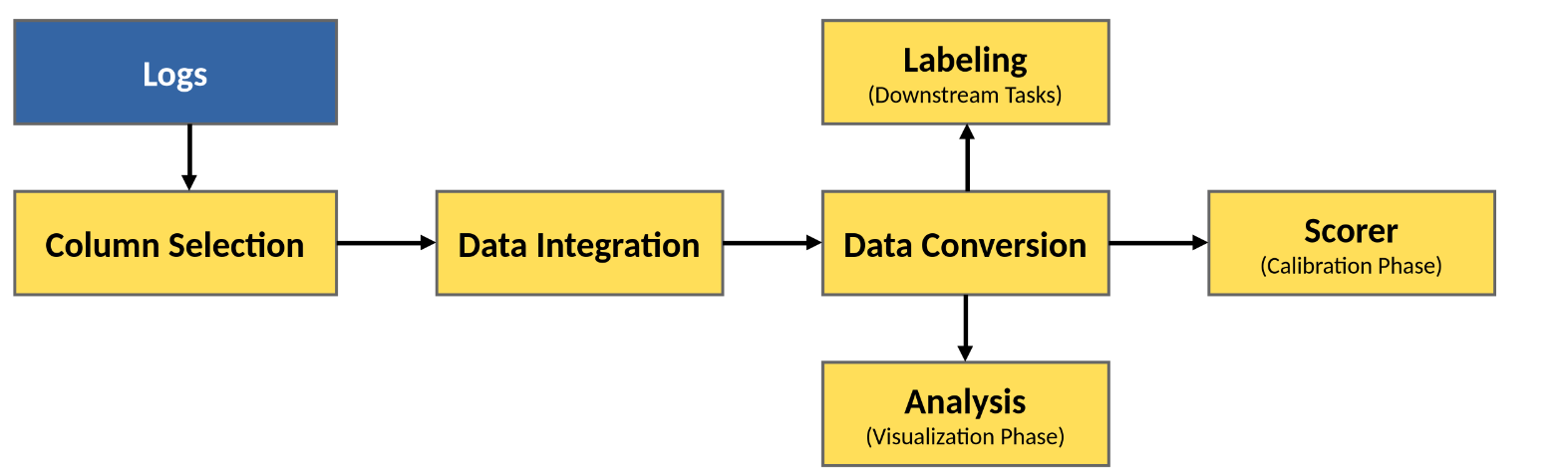}
    \caption{Post simulation log processing workflow (processing phase), illustrating column selection, data integration, conversion, and downstream usage for analysis, labeling, or calibration.}
    \label{fig:hdgen-pipeline-processing}
\end{figure}

The raw outputs generated by the simulator need be transformed into structured, application ready datasets which is done through a modular and scalable processing pipeline. After data generation, simulation logs can are processed in two primary ways: (1) by concatenating the complete set of split log files, or (2) by trimming and merging logs from a selected subset of simulation instances. In both cases, standardization is often required. For example, spatial coordinates recorded in the \texttt{agentStateTable} are not stored in standard GPS format and must be converted before downstream use.
Our processing scripts convert the raw logs into structured datasets suitable for tasks such as machine learning model training, calibration, and statistical analysis. The \texttt{agentStateTable}, which records the state of each agent at every simulation tick, can be split into manageable chunks, filtered by relevant fields, and reassembled into a unified dataset. Users can explicitly specify which fields are retained, enabling targeted data extraction tailored to specific research objectives.

As illustrated in \reffig{fig:hdgen-pipeline-processing}, post simulation processing begins with column selection, followed by integration of log fragments into a consolidated dataset. A data conversion step then transforms fields into formats required by downstream applications. The resulting dataset can be directly analyzed and visualized, labeled for supervised learning tasks, or passed to the scorer during the calibration phase without further modification.
HD-GEN is designed with scalability and customizability as core principles. To reduce unnecessary log generation, we provide detailed documentation on our Github page that guides users in configuring the simulator to emit only essential fields such as timestamps, geographic coordinates, and agent identifiers. This design minimizes storage and computational overhead while improving overall efficiency. For very large datasets, the pipeline includes Python scripts capable of processing hundreds of gigabytes of data, as well as lightweight Bash utilities that perform file system level concatenation and column extraction. In addition to filtering redundant fields, the pipeline supports embedding ground truth labels and other annotations, ensuring that generated datasets are well aligned with downstream research and application requirements.

    \subsubsection*{Visualization Phase (4)}
    \label{sec:visualization}
    Another critical aspect of HD-GEN is the visualization of processed data. Visualization offers a powerful means of interpreting simulation outputs by uncovering patterns, anomalies, and behavioral dynamics that are often difficult to detect in raw numerical logs \cite{grimm2002visual}. It is particularly valuable for validating the effects of parameter adjustments, identifying unexpected or emergent agent behaviors, and improving the overall accuracy of the model. When new features or anomalous behaviors arise, visual inspection provides an intuitive method for confirming their presence and assessing their impact across both temporal and spatial dimensions.
To facilitate these analyses, we provide a comprehensive visualization toolkit specifically designed for simulation logs. This toolkit supports a wide range of visualization modes, including time-series plots, spatial trajectory maps, and aggregated statistical summaries. By enabling researchers to rapidly explore simulation results, the toolkit enhances both debugging efficiency and interpretability of complex agent interactions. Together, these visualization capabilities complement the data processing pipeline, ensuring that raw outputs can be transformed not only into structured datasets but also into interpretable visual representations that accelerate validation and insight generation. Examples of the visualization output are shown in \reffig{fig:nhts-flows}--\reffig{fig:anomaly-example}

% \begin{figure}
%     \centering
%     \includegraphics[width=0.5\linewidth]{image.png}
%     \caption{Enter Caption}
%     \label{fig:placeholder}
% \end{figure}

% https://drive.google.com/file/d/18_uhJtCIOEHzsdB0loQFgymYNfjyyoVT/view
% https://par.nsf.gov/servlets/purl/10187146
% https://www.researchgate.net/profile/Hamdi-Kavak-2/publication/344737807_Advancing_Simulation_Experimentation_Capabilities_with_Runtime_Interventions/links/5ffddd2845851553a03d4b12/Advancing-Simulation-Experimentation-Capabilities-with-Runtime-Interventions.pdf
% https://www.osti.gov/servlets/purl/2476613

\section{Experimental Result}
\label{sec:results}
    In this section, we present the results of our experiments and analyses. We evaluate the realism of our simulations, analyze the impact of various parameters on the outcomes, and provide detailed simulation logs to demonstrate the performance and behavior of our system.
    
    \begin{table}[h]
        \centering
        \caption{Simulation scenarios used in the realism analysis.}
        \label{tab:realism-scenarios}
        \begin{tabular}{lrrl}
            \hline
            \textbf{Scenario} &
            \textbf{Number of agents} &
            \textbf{Duration (days)} &
            \textbf{Study area} \\
            \hline
            1k-360  & 1,000   & 360 & Atlanta \\
            10k-360 & 10,000  & 360 & Atlanta \\
            100k-75 & 100,000 & 75  & Atlanta \\
            \hline
        \end{tabular}
    \end{table}
    
    \subsection{Realism Analysis}
    \label{sec:realism}

\begin{figure}[t]
\centering

\begin{minipage}{0.49\textwidth}
    \centering
    \includegraphics[
        width=\linewidth, 
        trim=1.5cm 1.5cm 1.5cm 4cm, clip
    ]{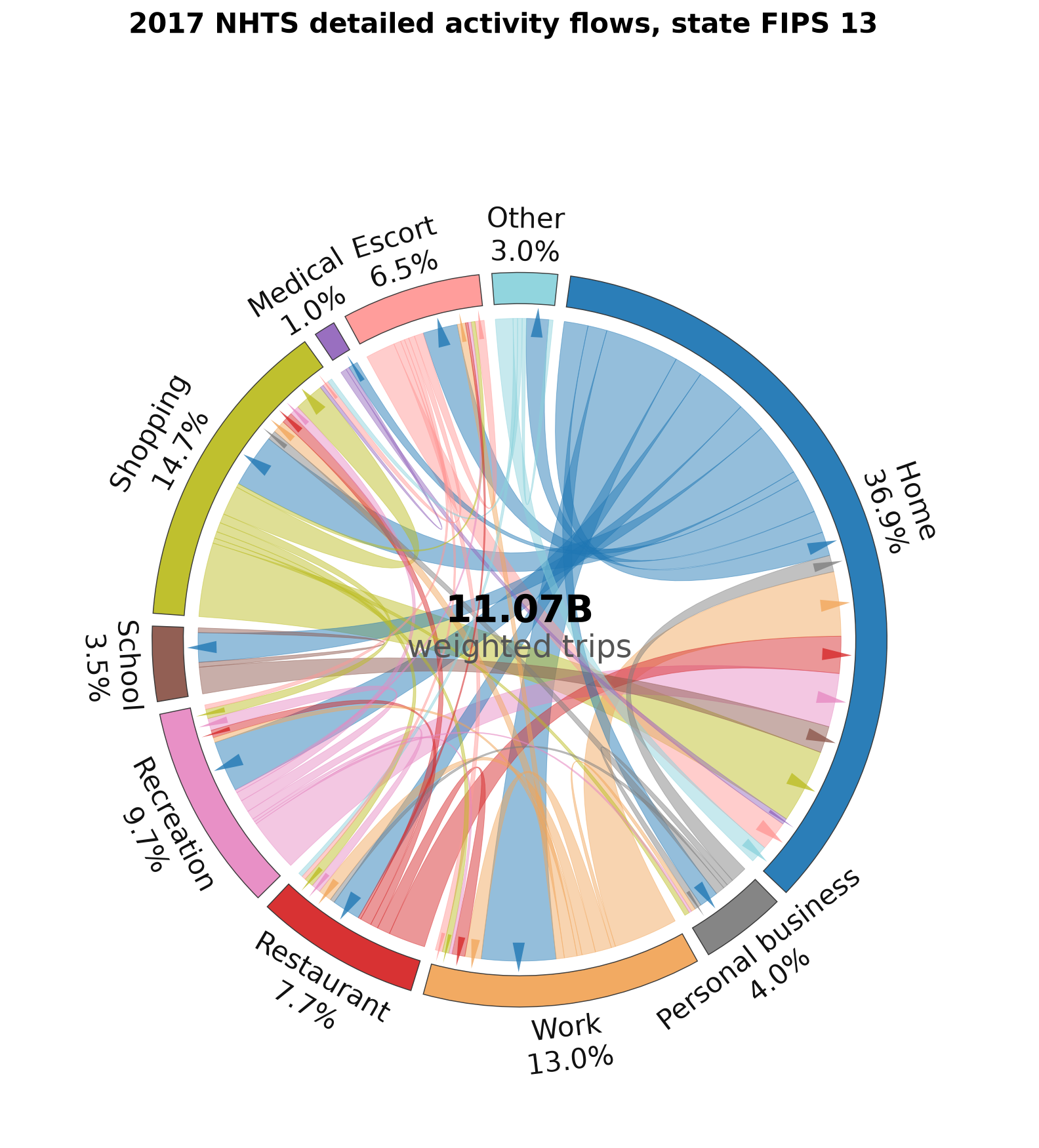}
    \par
    \small (a) All NHTS 2017 destination categories.
\end{minipage}
\hfill
\begin{minipage}{0.49\textwidth}
    \centering
    \includegraphics[
        width=\linewidth,
        trim=1.5cm 1.5cm 1.5cm 4cm, clip
    ]{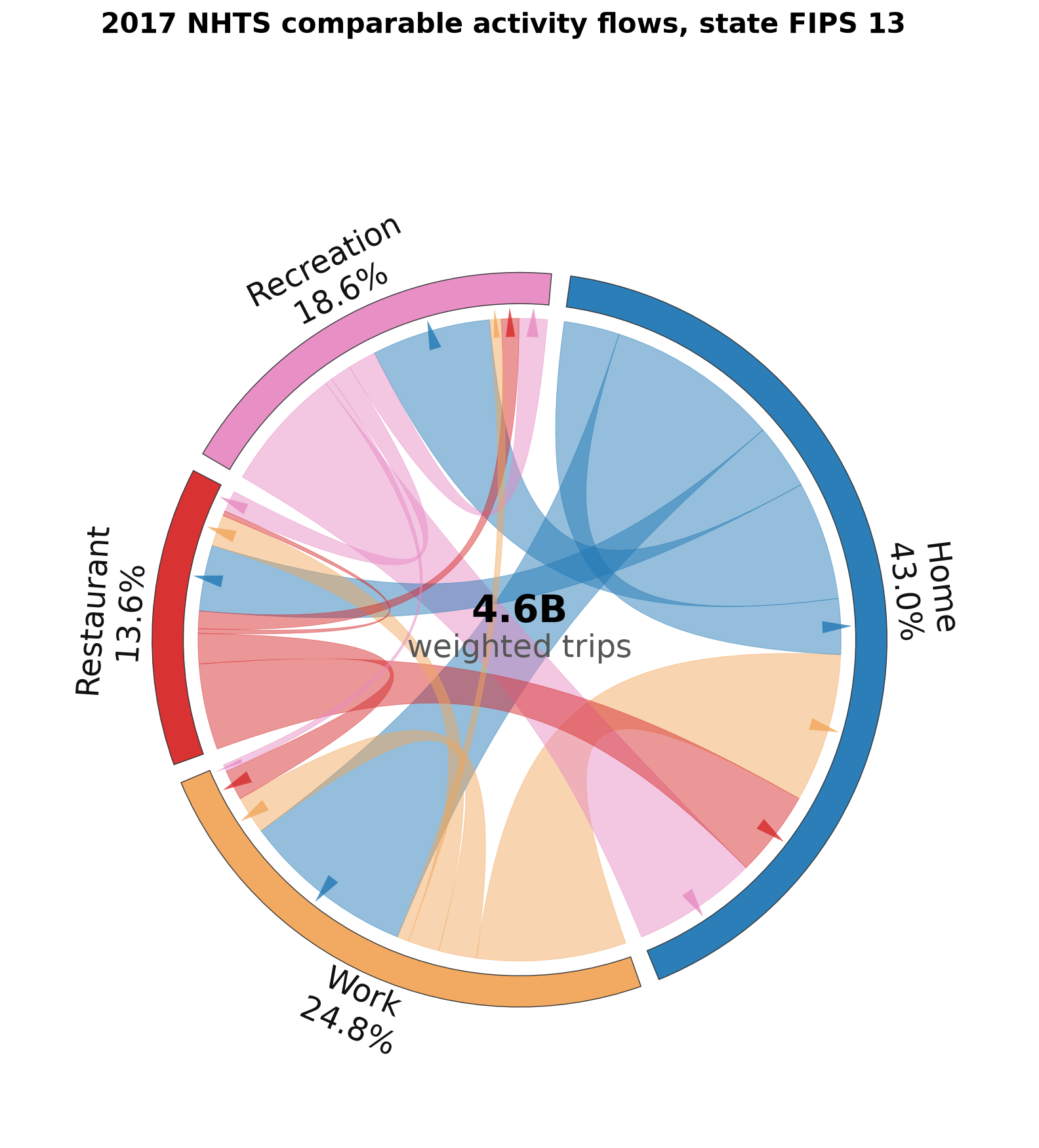}
    \par
    \small (b) NHTS 2017 categories represented in the simulation.
\end{minipage}

\caption{Source--destination travel flows obtained from the Georgia NHTS 2017 data. Ribbon color represents the source category, and the arrowhead represents the destination category.}
\label{fig:nhts-flows}
\end{figure}

To evaluate the realism of the simulated mobility patterns, we conducted three simulation scenarios with different population sizes and durations. The scenarios are summarized in~\reftab{tab:realism-scenarios}. The 1k-360 and 10k-360 scenarios simulate 1,000 and 10,000 agents, respectively, for a full 360-day period. The 100k-75 scenario simulates 100,000 agents for 75 days. All scenarios use a map of Atlanta, Georgia, USA.
The results of these scenarios are compared with travel patterns obtained from the 2017 National Household Travel Survey (NHTS 2017)~\cite{nhts2017}. The comparison focuses on the distribution of trips among different destination types and the flows between those destinations.
The simulation represents four place categories: Home, Work, Restaurant, and Recreation. NHTS 2017, however, provides a more detailed classification that also includes Shopping, School, Medical, Personal Business, Escort, and Other destinations. To illustrate both the complete NHTS travel structure and the portion that is directly comparable with the simulation,~\reffig{fig:nhts-flows} presents two source--destination chord diagrams for the state of Georgia.
In each diagram, the width of an outer segment represents the proportion of weighted trips associated with a destination category. The width of each ribbon represents the number of weighted trips between two categories. Ribbon color identifies the source category, while the arrowhead identifies the destination category.

\begin{figure}[t]
\centering

\begin{minipage}{0.32\textwidth}
    \centering
    \includegraphics[width=\linewidth,trim=1.5cm 1.5cm 1.5cm 4cm, clip]{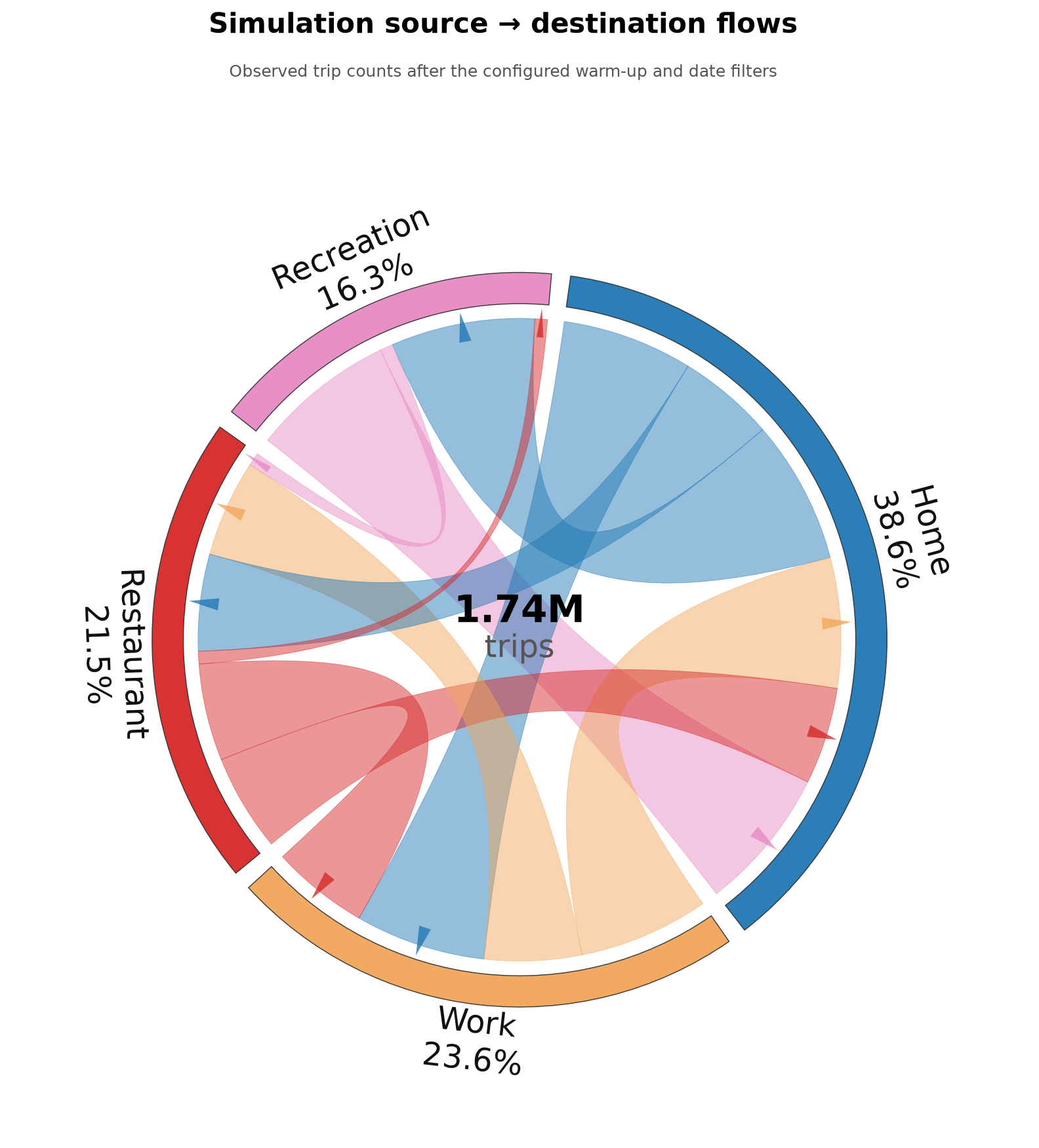}
    \par
    \small (a) Scenario 1: 1k-360.
\end{minipage}
\hfill
\begin{minipage}{0.32\textwidth}
    \centering
    \includegraphics[width=\linewidth,trim=1.5cm 1.5cm 1.5cm 4cm, clip]{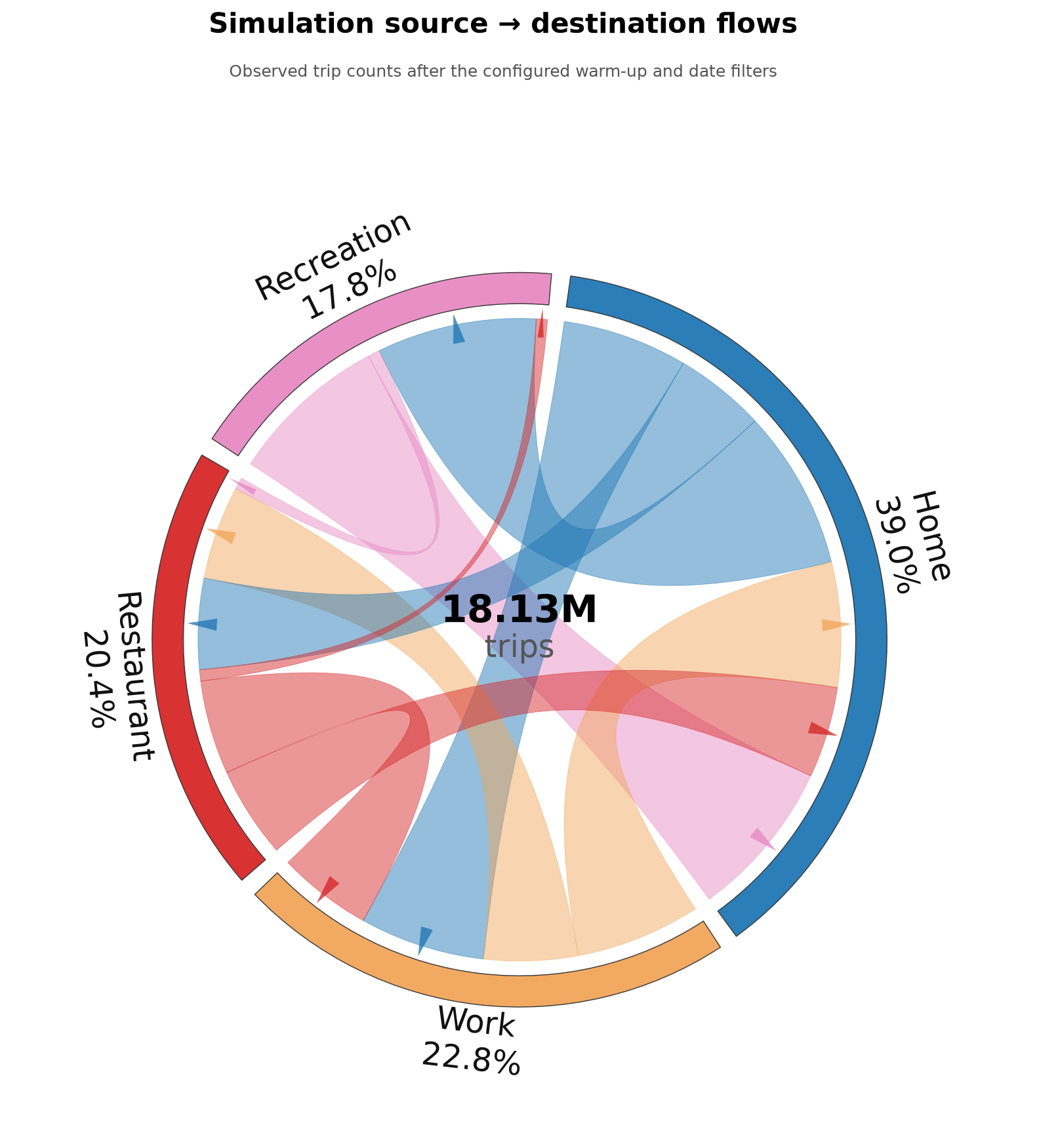}
    \par
    \small (b) Scenario 2: 10k-360.
\end{minipage}
\hfill
\begin{minipage}{0.32\textwidth}
    \centering
    \includegraphics[width=\linewidth,trim=1.5cm 1.5cm 1.5cm 4cm, clip]{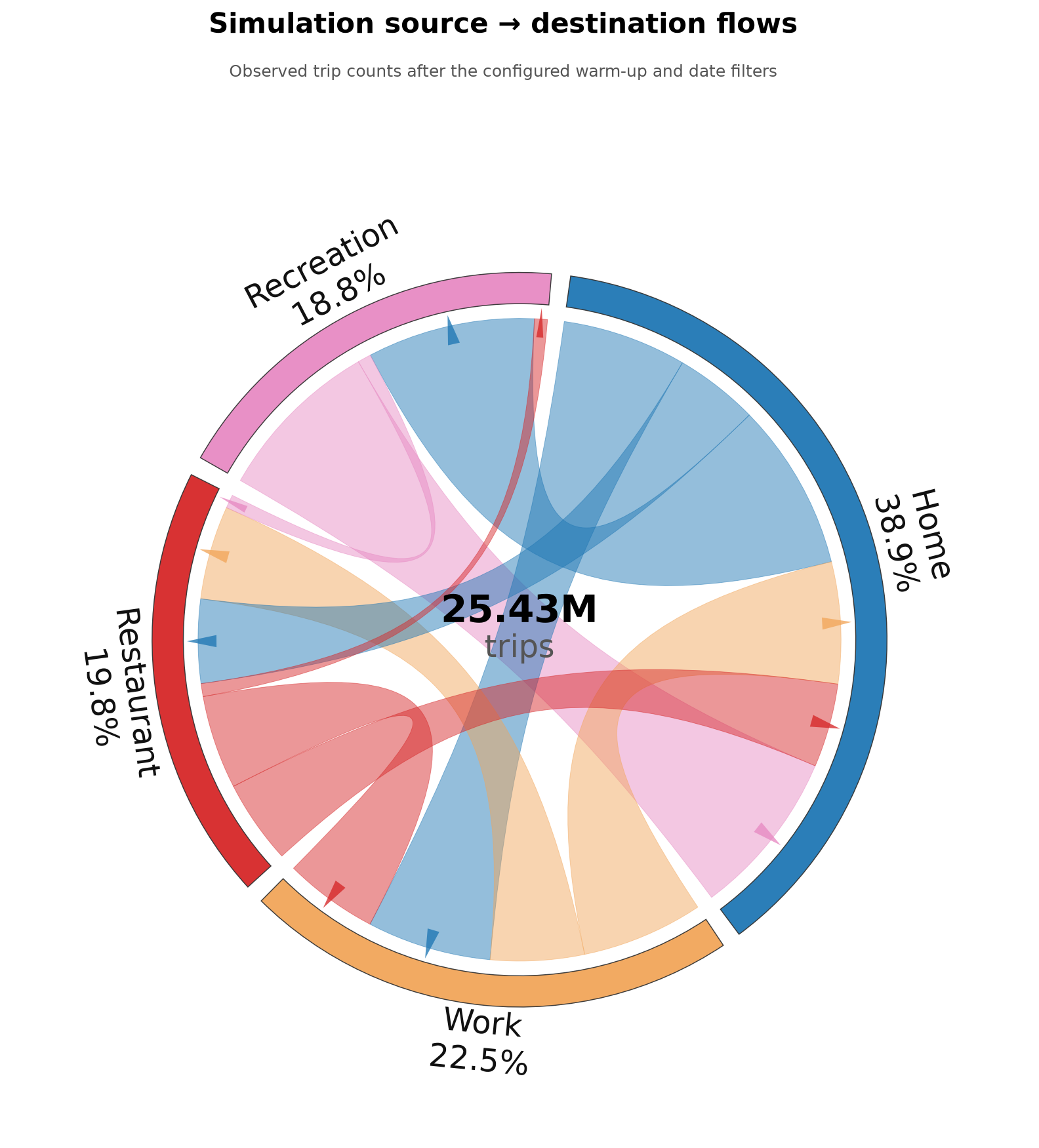}
    \par
    \small (c) Scenario 3: 100k-75.
\end{minipage}

\caption{Simulation source--destination flows after removing the 30-day warm-up period. Ribbon color indicates the source category, and the arrowhead indicates the destination category.}
\label{fig:simulation-flows}

\end{figure}

\reffig{fig:nhts-flows}(a) shows approximately (11.07) billion weighted trips across all NHTS 2017 categories. Home is the largest category, accounting for (36.9\%) of the trips. It is followed by Shopping at (14.7\%), Work at (13.0\%), Recreation at (9.7\%), Restaurant at (7.7\%), Escort at (6.5\%), Personal Business at (4.0\%), School at (3.5\%), Medical at (1.0\%), and Other at (0.2\%).
The large number and width of the ribbons connected to Home indicate that home is the primary origin or destination of daily travel. Substantial flows are also observed between Home and Work, Home and Shopping, and Home and Recreation. The detailed diagram further demonstrates that real-world travel behavior includes many trip purposes that are not explicitly represented in the simulation.
\reffig{fig:nhts-flows}(b) restricts the NHTS 2017 data to the four categories represented by the simulation. After this filtering, approximately (4.6) billion weighted trips remain. Within this reduced set, Home accounts for (43.0\%) of the trips, Work for (24.8\%), Recreation for (18.6\%), and Restaurant for (13.6\%).
The percentages in~\reffig{fig:nhts-flows}(b) are larger than their corresponding percentages in~\reffig{fig:nhts-flows}(a) because they are normalized over only the four retained categories. Home remains the dominant category, and the strongest movements are associated with trips between Home and Work and between Home and Recreation. This reduced NHTS representation provides a consistent reference for evaluating whether the three simulation scenarios reproduce realistic source--destination travel patterns.
The source--destination flows of the three simulation scenarios are shown in~\reffig{fig:simulation-flows}. These figures are computed after removing the first 30 days of each simulation. The first 30 days are treated as a warm-up period because agents are still establishing their routines and satisfying initial basic needs. During this period, their behavior can be atypical and may not represent stable long-term mobility patterns. Excluding these early days allows the analysis to focus on the steady-state behavior of the simulated population.

\reffig{fig:simulation-flows}(a) corresponds to Scenario 1 (1k-360) and contains approximately (1.74) million trips. Home is the dominant category with (38.6\%) of trips, followed by Work ((23.6\%)), Restaurant ((21.5\%)), and Recreation ((16.3\%)). The strongest flows are centered around Home, indicating that most trips either start from or end at home. Large exchanges are also visible between Home and Work and between Home and Restaurant.
\reffig{fig:simulation-flows}(b) shows Scenario 2 (10k-360), with approximately (18.13) million trips. The destination shares are Home ((39.0\%)), Work ((22.8\%)), Restaurant ((20.4\%)), and Recreation ((17.8\%)). The overall structure is very similar to Scenario 1, which suggests that the simulation behavior is stable as the population increases from 1,000 to 10,000 agents. Compared with Scenario 1, the Restaurant share decreases slightly, while Recreation increases slightly.
\reffig{fig:simulation-flows}(c) presents Scenario 3 (100k-75), which contains approximately (25.43) million trips. The category shares are Home ((38.9\%)), Work ((22.5\%)), Restaurant ((19.8\%)), and Recreation ((18.8\%)). Again, Home remains the largest destination category. Relative to the first two scenarios, the distribution becomes slightly more balanced, with a continued reduction in the Restaurant share and a continued increase in the Recreation share.

Across all three scenarios, the overall pattern is consistent. Home is always the largest category, accounting for about (39\%) of trips, while Work remains the second largest category at about (22\%) to (24\%). Restaurant and Recreation together account for the remaining trips, with Restaurant gradually decreasing and Recreation gradually increasing as the scenario size grows. This consistency indicates that the simulation produces stable travel patterns across different population sizes and durations.
When compared with the reduced NHTS 2017 distribution for the same four categories, the simulation reproduces some important qualitative features, especially the dominant role of Home and the strong interaction between Home and Work. However, the simulation allocates a larger share of trips to Restaurant and a smaller share to Recreation than observed in NHTS 2017. Recreation is very close to the empirical benchmark, while Work is somewhat lower. These differences suggest that the simulation captures the broad structure of daily mobility, but some activity preferences, especially those related to Restaurant and Home, may still require further calibration.

\begin{figure}[t]
    \centering
    \includegraphics[width=\linewidth]{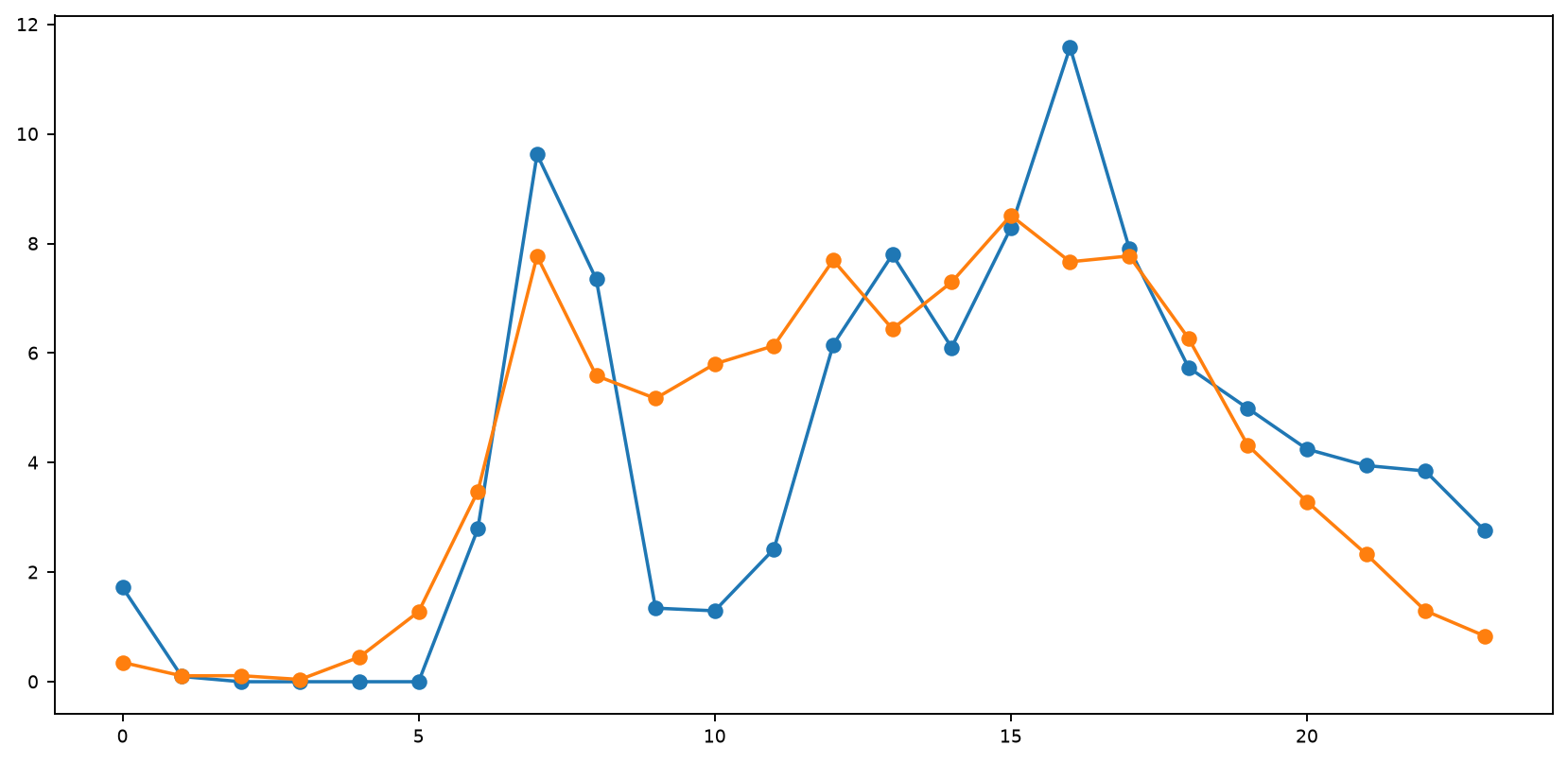}
    \caption{Comparison of the hourly departure-time distributions produced by the simulation and reported in NHTS 2017. The simulation captures the general daily travel pattern and produces distinct morning and afternoon commuting peaks.}
    \label{fig:departure-time-comparison}
\end{figure}

\reffig{fig:departure-time-comparison} compares the hourly departure-time distributions produced by the simulation with those reported in NHTS 2017. Both profiles show relatively little travel during the early morning, an increase beginning around 6:00, substantial travel throughout the daytime, and a decline during the evening.
The simulation produces two distinct rush-hour peaks. The morning peak occurs around 7:00, when approximately (9.8\%) of simulated trips depart, while the largest afternoon peak occurs around 16:00 and accounts for approximately (11.8\%) of trips. These peaks reflect the movement of agents to and from work and demonstrate that the simulation reproduces the expected temporal effects of daily commuting.
The NHTS profile is smoother and less concentrated. It shows a morning increase around 7:00, followed by a relatively broad distribution of departures between approximately 10:00 and 18:00. Its afternoon peak occurs around 15:00--16:00, but it is considerably smaller than the corresponding simulation peak. It's interesting to note that the NHTS data does not exhibit the typical morning-peak and evening-peak observed during rush hours which is certainly observable in the real-world in the Atlanta region. 
This difference can be explained because NHTS includes a wider range of trip purposes and individual schedules, beyond home-work-home commute which distribute departures more evenly throughout the day but which are not modeled in the simulation.

Overall, the simulation captures the main temporal characteristics of real-world mobility, including low overnight activity, increased morning travel, sustained daytime activity, an afternoon commuting peak, and declining evening travel. However, the simulated peaks are sharper than those observed in NHTS, and the simulation underestimates departures during the late morning. Therefore, the results indicate that the simulation provides a broadly representative daily travel profile, while additional calibration could improve the distribution of midday trips and reduce the concentration of departures during commuting hours.

    \subsection{Parameters Analysis}
    \label{sec:parameters}

This section examines how the sampled parameters are associated with simulation completion, execution time, and mobility outcomes. Two configurations are considered: Vanilla and HD-GEN with each 1,000 parameterized simulation instances were evaluated for each configuration. Parameter effects are measured using Spearman's rank correlation, denoted by $\rho$. We additionally compare the mean outcome in the upper and lower parameter quintiles. These results describe associations rather than causal effects, because several parameters vary simultaneously in each instance.

\begin{figure}[t]
    \centering

    \begin{minipage}[t]{0.49\textwidth}
        \centering
        \includegraphics[width=\linewidth]{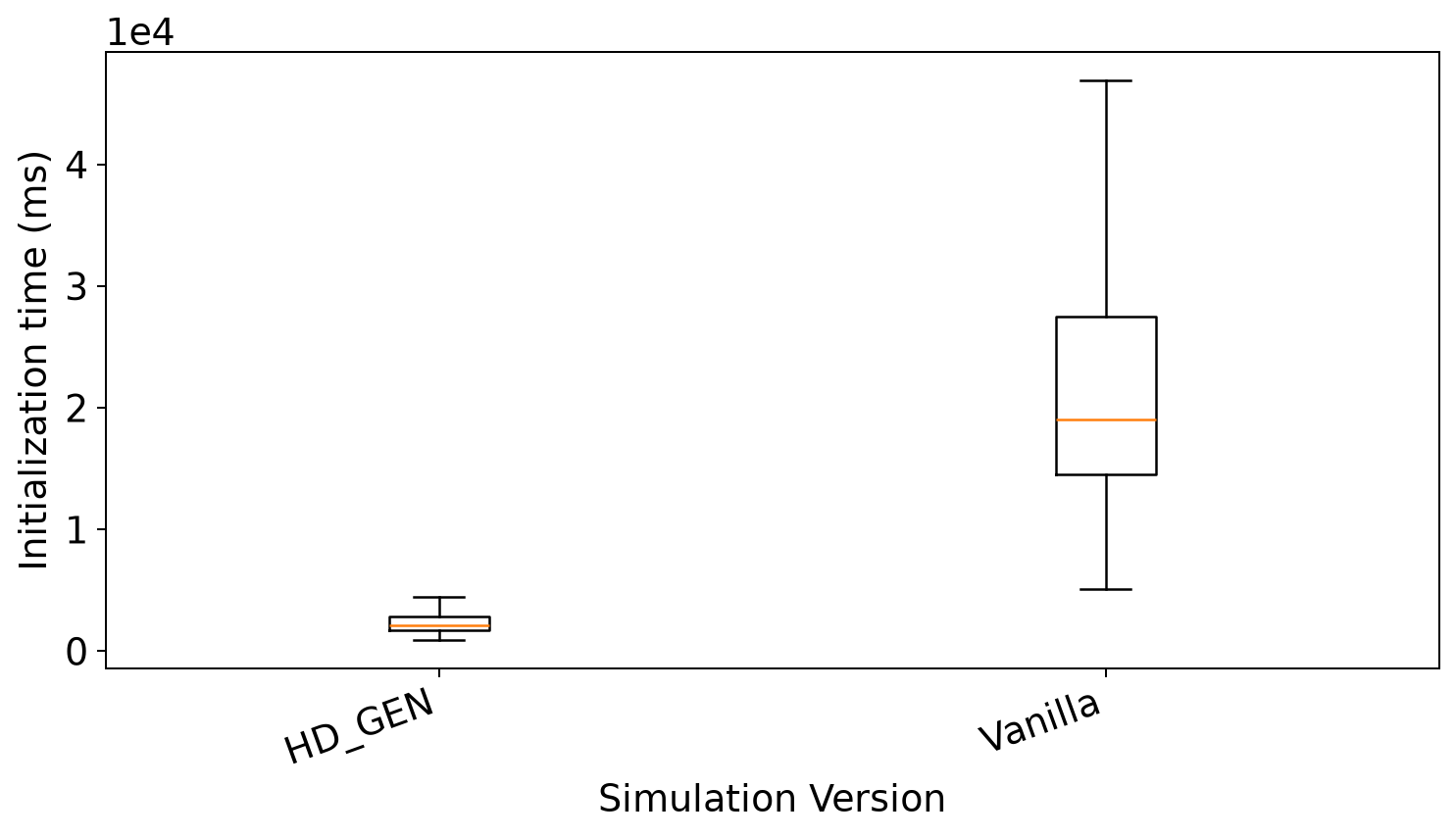}
        \par\smallskip
        \textbf{(a)} Initialization time.
    \end{minipage}
    \hfill
    \begin{minipage}[t]{0.49\textwidth}
        \centering
        \includegraphics[width=\linewidth]{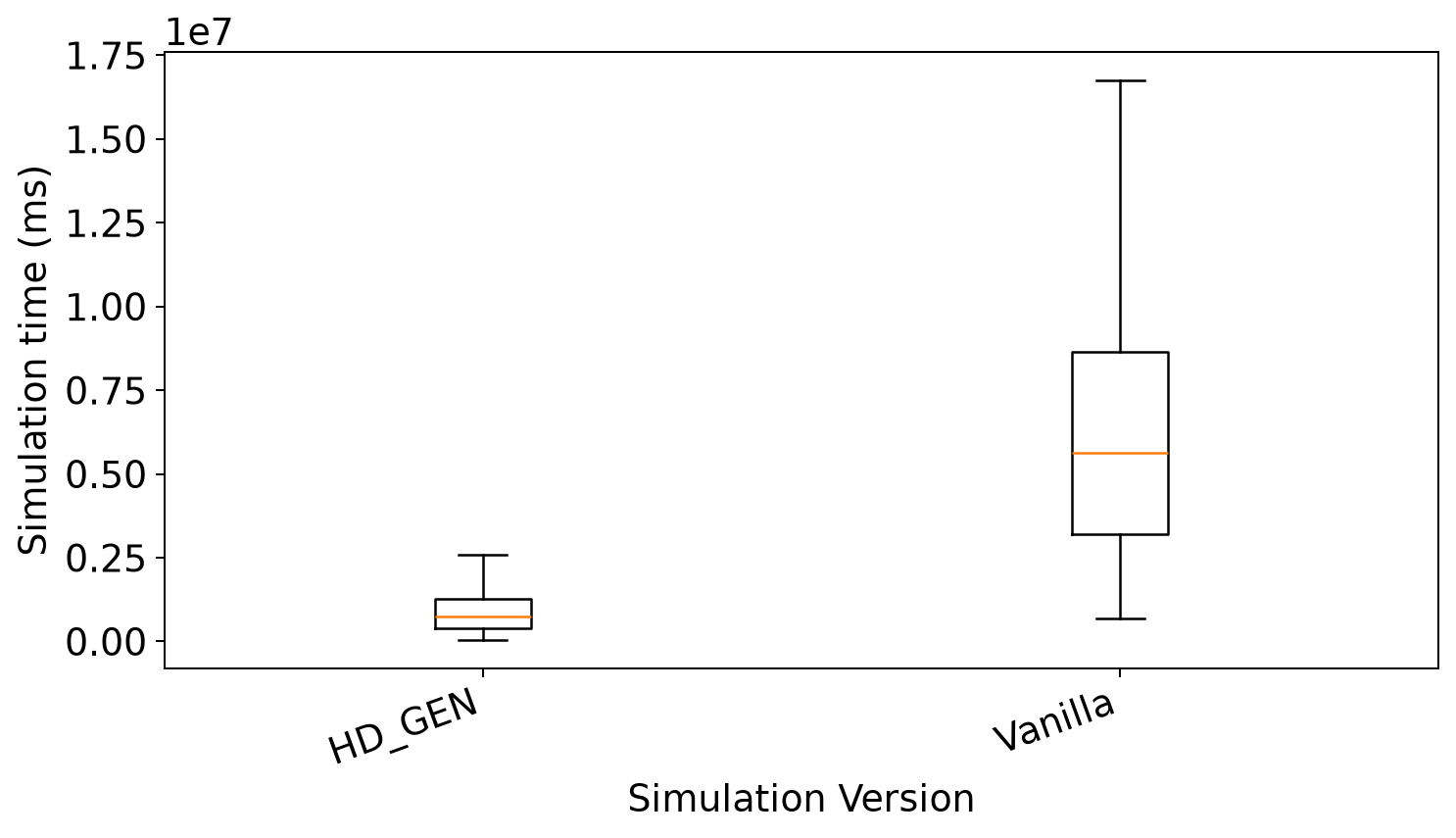}
        \par\smallskip
        \textbf{(b)} Simulation time.
    \end{minipage}

    \caption{Comparison of computational performance between the HD\_GEN and Vanilla scenarios. The box plots show the distributions of (a) initialization time and (b) simulation time in milliseconds.}
    \label{fig:runtime_comparison}
\end{figure}

\reffig{fig:runtime_comparison} compares initialization and simulation times for the HD\_GEN and Vanilla scenarios. HD\_GEN consistently requires substantially less computational time than Vanilla for both metrics. The median initialization time of HD\_GEN is approximately \(2{,}000\) ms, compared with roughly \(19{,}000\) ms for Vanilla, while its median simulation time is below \(1\times10^{6}\) ms compared with approximately \(5.5\times10^{6}\) ms for Vanilla. The narrower boxes and shorter whiskers for HD\_GEN also indicate lower runtime variability, whereas Vanilla exhibits a much wider distribution and considerably higher maximum values. Overall, these results show that HD\_GEN provides faster and more predictable initialization and simulation performance.

\begin{figure}[t]
    \centering

    \begin{minipage}[t]{0.49\textwidth}
        \centering
        \includegraphics[width=\linewidth]
        {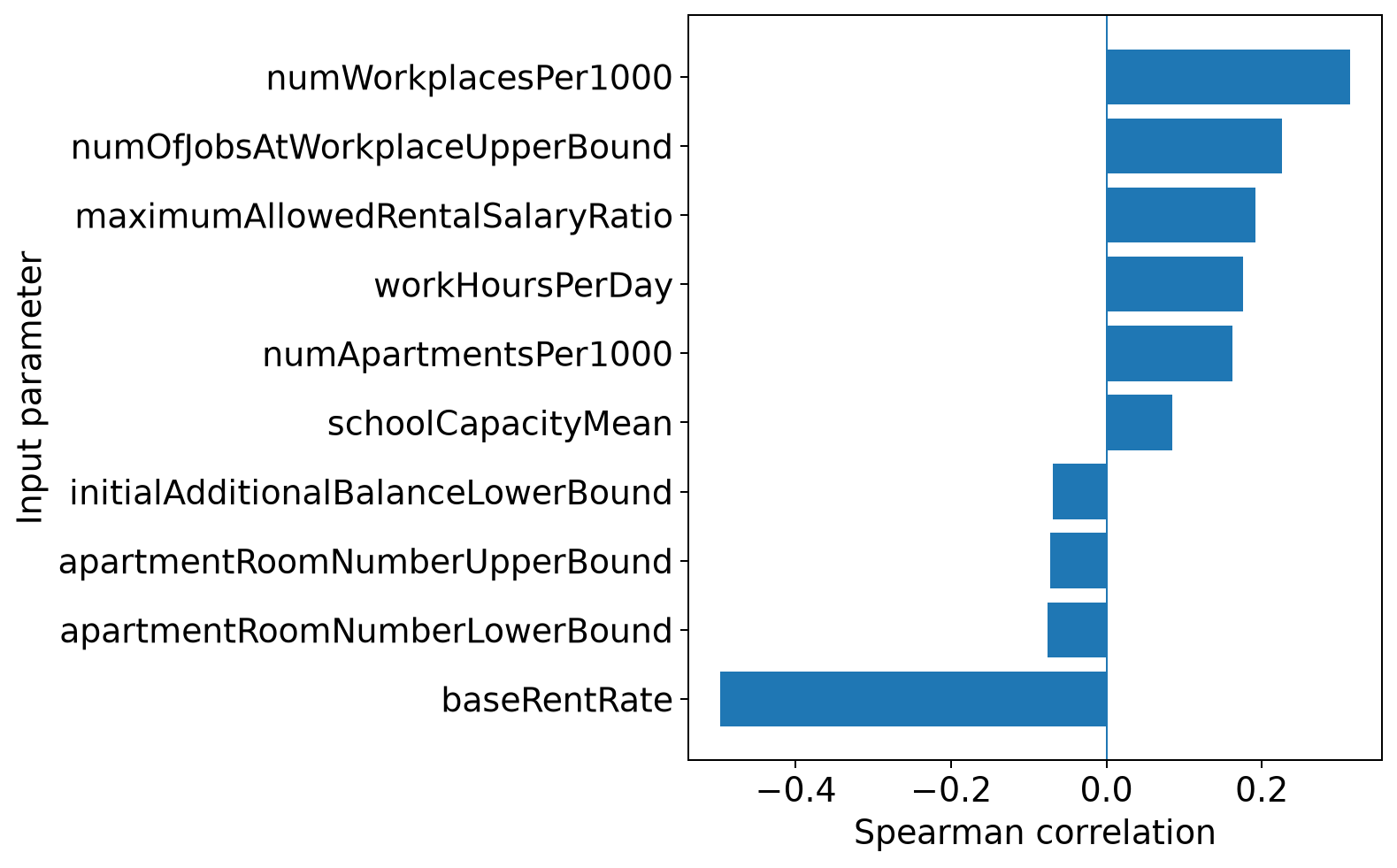}
        \caption*{(a) Initialization-time correlations.}
    \end{minipage}
    \hfill
    \begin{minipage}[t]{0.49\textwidth}
        \centering
        \includegraphics[width=\linewidth]
        {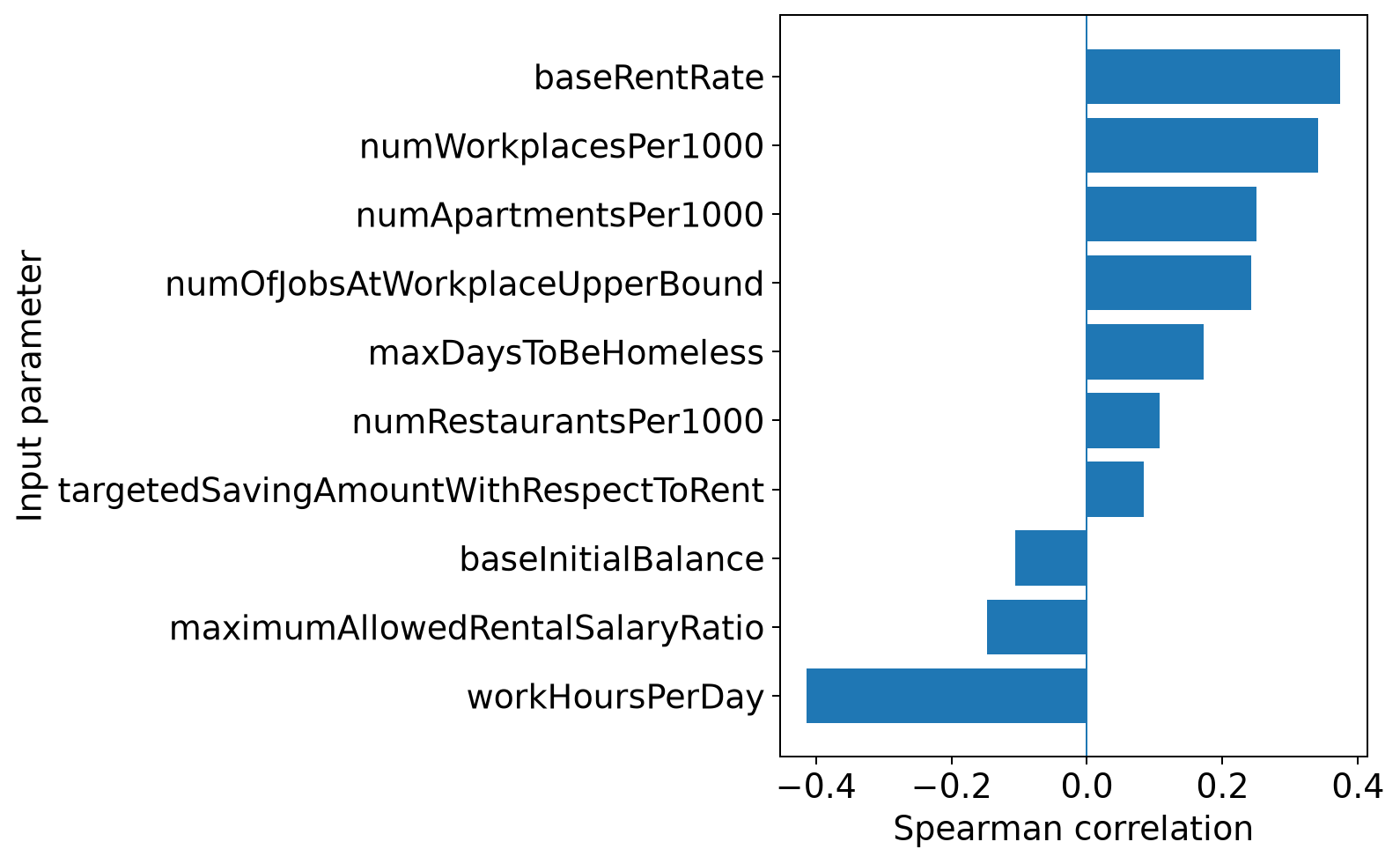}
        \caption*{(b) Simulation-time correlations.}
    \end{minipage}

    \caption{Strongest Spearman correlations between input parameters and runtime metrics for the Vanilla scenario. Positive values indicate that larger parameter values are associated with longer runtimes, while negative values indicate inverse associations.}
    \label{fig:vanilla_runtime_correlations}
\end{figure}

\reffig{fig:vanilla_runtime_correlations} shows the strongest Spearman correlations between model inputs and runtime for the Vanilla scenario. Initialization time is most strongly and negatively associated with \texttt{baseRentRate} ($\rho \approx -0.50$), while \texttt{numWorkplacesPer1000} has the largest positive association ($\rho \approx 0.31$); several other workplace, housing, and labor parameters exhibit weaker positive relationships. In contrast, simulation time is most positively associated with \texttt{baseRentRate} ($\rho \approx 0.37$) and \texttt{numWorkplacesPer1000} ($\rho \approx 0.34$), whereas \texttt{workHoursPerDay} shows the strongest negative association ($\rho \approx -0.42$). Overall, the results indicate that the factors affecting initialization and simulation time differ substantially, and that the observed relationships are moderate rather than strongly deterministic.

\begin{figure}[t]
    \centering

    \begin{minipage}[t]{0.49\textwidth}
        \centering
        \includegraphics[width=\linewidth]
        {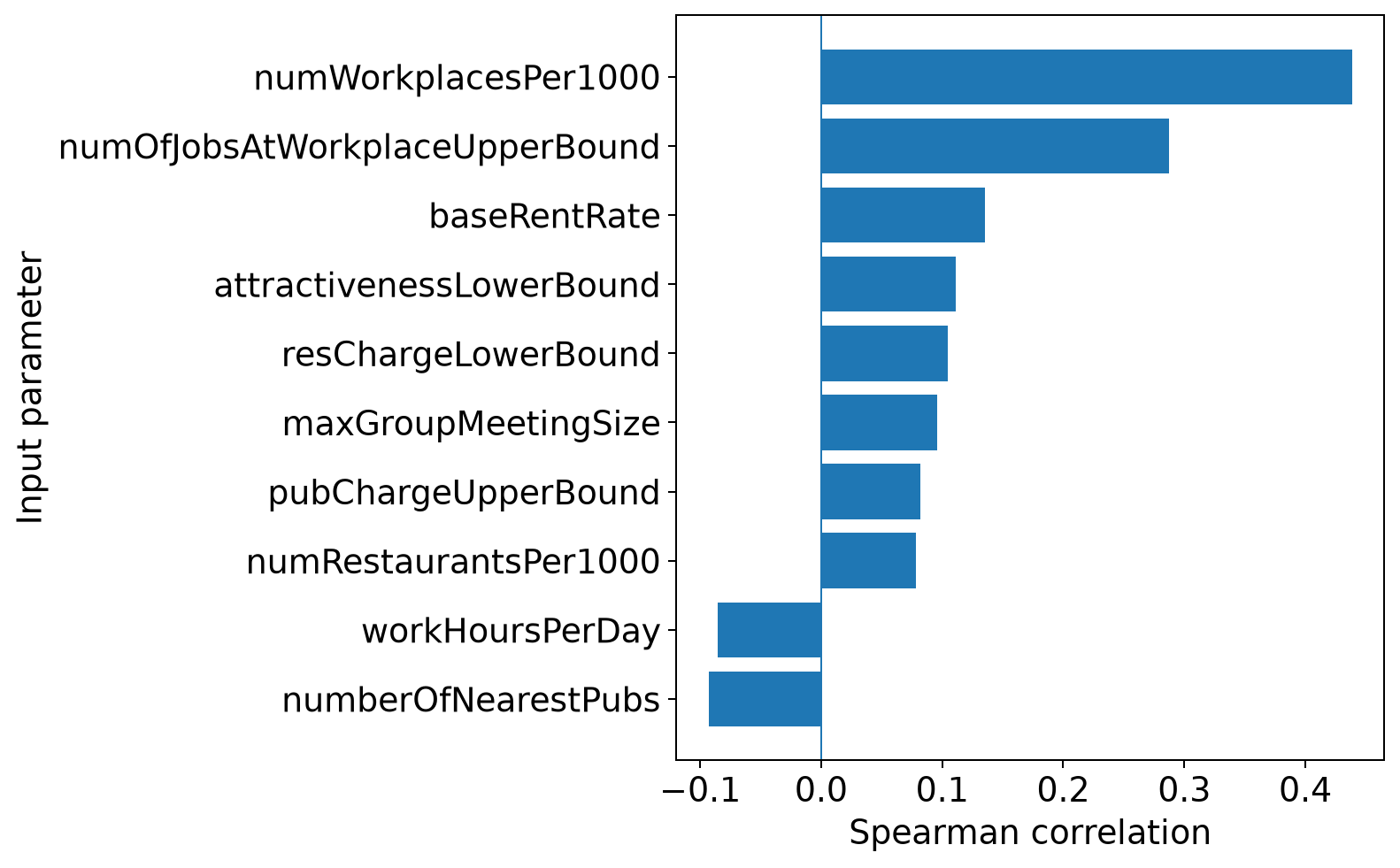}
        \caption*{(a) Initialization-time correlations.}
    \end{minipage}
    \hfill
    \begin{minipage}[t]{0.49\textwidth}
        \centering
        \includegraphics[width=\linewidth]
        {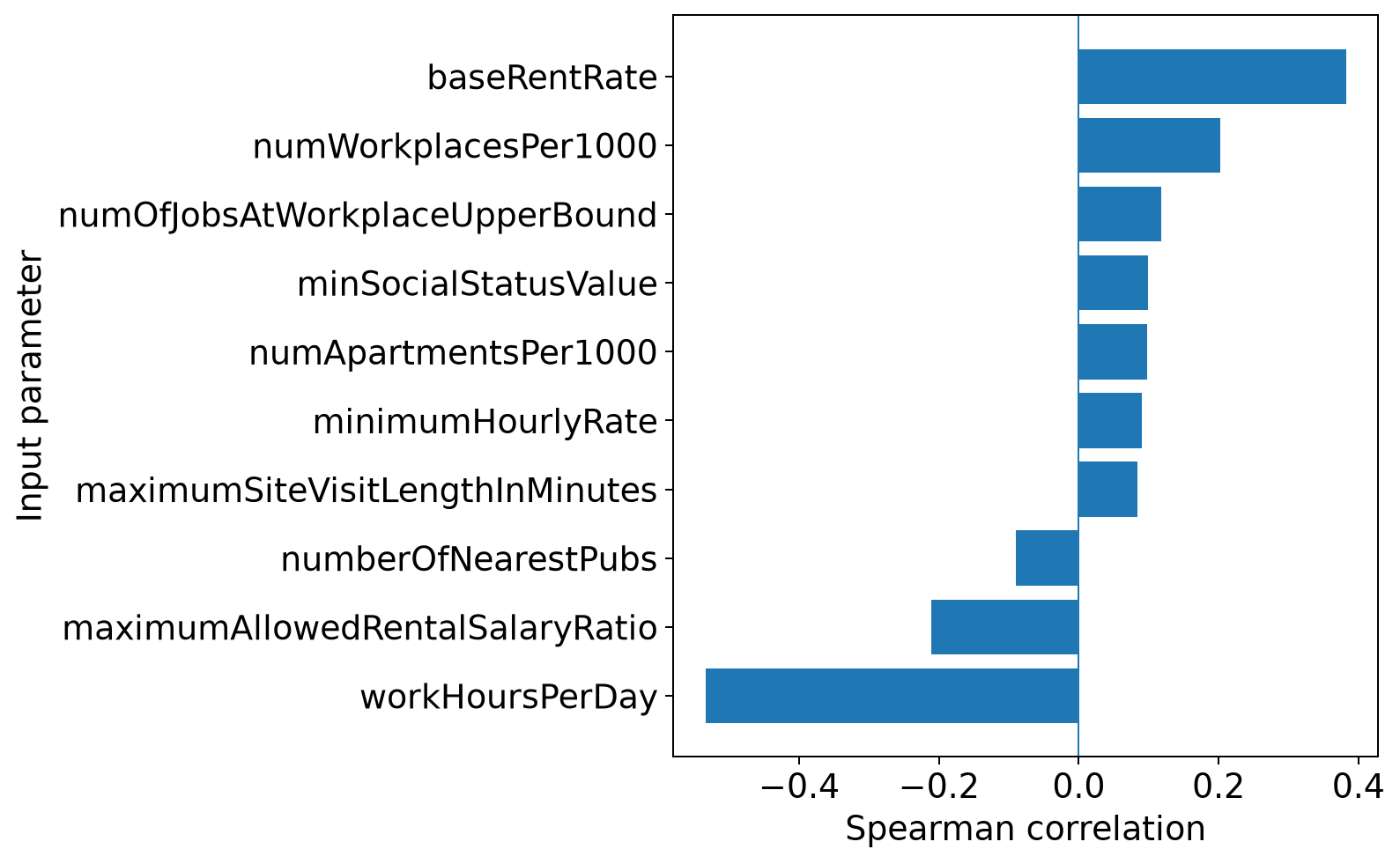}
        \caption*{(b) Simulation-time correlations.}
    \end{minipage}

    \caption{Strongest Spearman correlations between HD\_GEN input parameters and runtime metrics. Positive values indicate that larger parameter values are associated with longer runtimes, whereas negative values indicate inverse associations.}
    \label{fig:hdgen_runtime_correlations}
\end{figure}

\reffig{fig:hdgen_runtime_correlations} presents the strongest Spearman correlations between the HD\_GEN input parameters and its runtime metrics. Initialization time is most positively associated with \texttt{numWorkplacesPer1000} ($\rho \approx 0.44$) and \texttt{numOfJobsAtWorkplaceUpperBound} ($\rho \approx 0.29$), while \texttt{workHoursPerDay} and \texttt{numberOfNearestPubs} show relatively weak negative associations. Simulation time is most positively associated with \texttt{baseRentRate} ($\rho \approx 0.38$), followed by \texttt{numWorkplacesPer1000} ($\rho \approx 0.20$), whereas \texttt{workHoursPerDay} has the strongest negative relationship ($\rho \approx -0.53$) and \texttt{maximumAllowedRentalSalaryRatio} also shows a moderate negative association ($\rho \approx -0.21$). Overall, workplace availability appears to increase both initialization and simulation costs, while longer configured work hours are associated with shorter simulation runtimes; these correlations describe monotonic associations and should not be interpreted as causal effects.

We further analysed the raw data, and the results are summarised as follows.
The rent parameter, \texttt{baseRentRate}, had the broadest association with the output. It was negatively correlated with initialization time ($\rho=-0.488$), number of check-ins ($\rho=-0.288$), number of trips ($\rho=-0.289$), and average travel distance ($\rho=-0.271$). Relative to the lower quintile, the upper rent-rate quintile had 77.2\% lower mean initialization time, 29.9\% fewer trips, and 25.9\% lower average distance.
The number of workplaces, \texttt{numWorkplacesPer1000}, was the dominant parameter for spatial coverage. Its correlation with the number of unique visited locations was $\rho=0.704$, and the upper quintile produced 18.6\% more unique locations than the lower quintile. It was also positively associated with initialization and simulation time, indicating that a denser workplace environment increases both the explored location set and the computational workload. In contrast, \texttt{workHoursPerDay} was negatively correlated with simulation time ($\rho=-0.414$); the upper quintile reduced mean simulation time by 51.4\%.
Several effects were consistent across configurations. First, \texttt{numWorkplacesPer1000} strongly increased spatial coverage in both Vanilla modes and in HDGen--All. Second, \texttt{workHoursPerDay} was the dominant simulation-time parameter in different configurations, with higher values associated with shorter execution times. Finally, \texttt{baseRentRate} consistently reduced activity volume or travel distance, particularly in the Vanilla experiments.

    \subsection{Simulation Results}
    \label{sec:logs}
    % \begin{figure}[htbp]
% \centering
% \includegraphics[width=\linewidth]{figs/education_distribution.png}
% \caption{Distribution of agents by highest education level.}
% \label{fig:education-distribution}
% \end{figure}

% \reffig{fig:education-distribution} presents the education distribution of the simulated population. The largest group consists of agents classified as having a high-school or college-level education, representing approximately 54 agents. This is followed by agents with bachelor's degrees, with approximately 23 agents. Graduate-level and low-education categories contain approximately 12 and 10 agents, respectively. The distribution indicates that most agents have an intermediate education level, while smaller proportions are assigned to the highest and lowest education categories.

This section presents selected results and insights from the 1k-360 scenario. The complete results for this scenario, together with those for the 10k-360 and 100k-75 scenarios, are available in the project’s GitHub repository.

\begin{figure}[htbp]
\centering
\includegraphics[width=\linewidth,trim=0cm 0cm 0cm 0.8cm, clip]{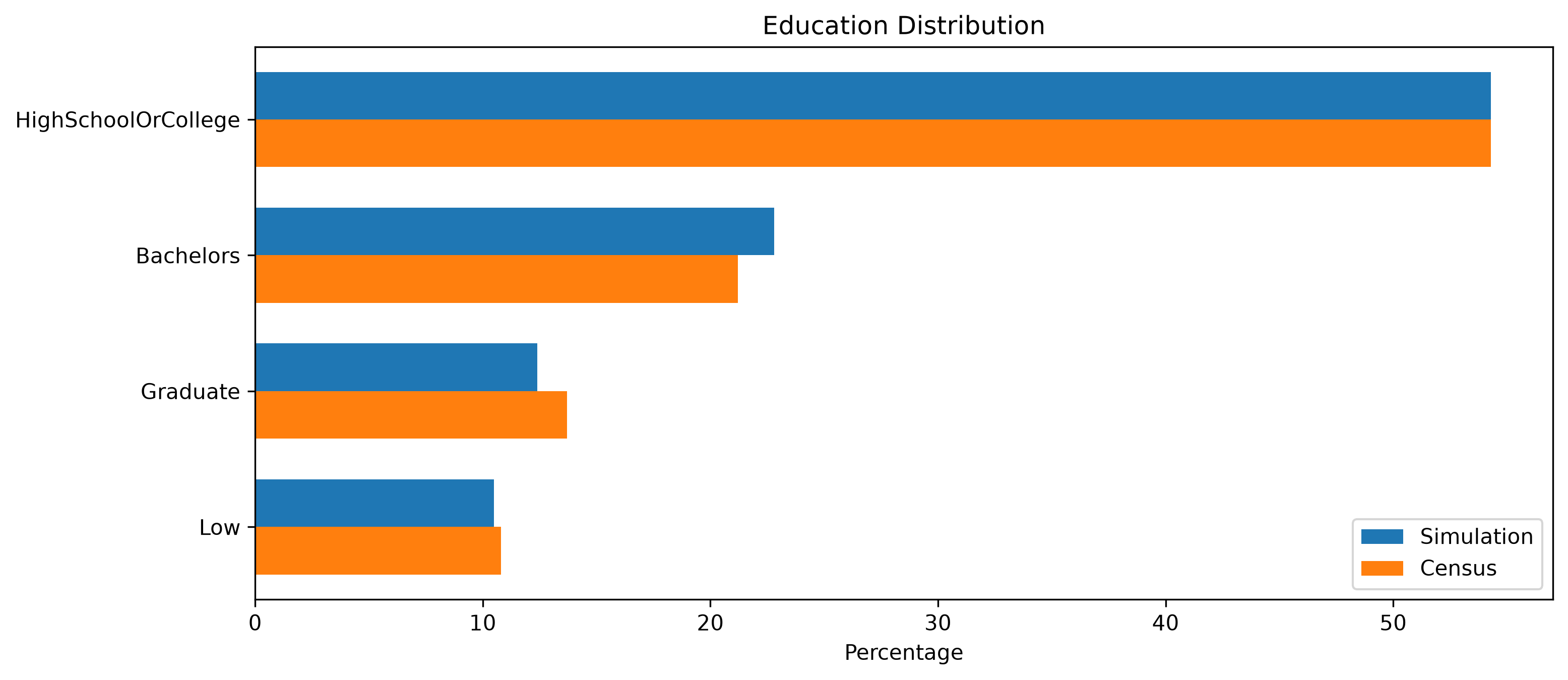}
\caption{Comparison of education-level distributions in the simulated population and Census data.}
\label{fig:education-distribution}
\end{figure}

\reffig{fig:education-distribution} compares the education distribution of the simulated population with the corresponding Census distribution. In both datasets, the largest category is HighSchoolOrCollege, accounting for approximately 54\% of the population. The simulation contains a slightly higher proportion of agents with bachelor's degrees, approximately 23\% compared with 21\% in the Census data. Graduate-level education accounts for approximately 12.5\% of the simulated population and 13.5\% of the Census population, while the low-education category represents approximately 10.5\% in both distributions. Overall, the simulated distribution closely reproduces the Census benchmarks, with only minor differences across education categories. Education proportions are configurable simulation parameters and, by default, are based on the 2009 U.S. Census distribution, which provides a reasonable approximation for Georgia. Each agent's education level is then sampled randomly according to these proportions, although the default values can be replaced with either Georgia-specific or national data.

\begin{figure}[htbp]
\centering
\includegraphics[width=0.8\linewidth,trim=0cm 0cm 0cm 0.7cm, clip]{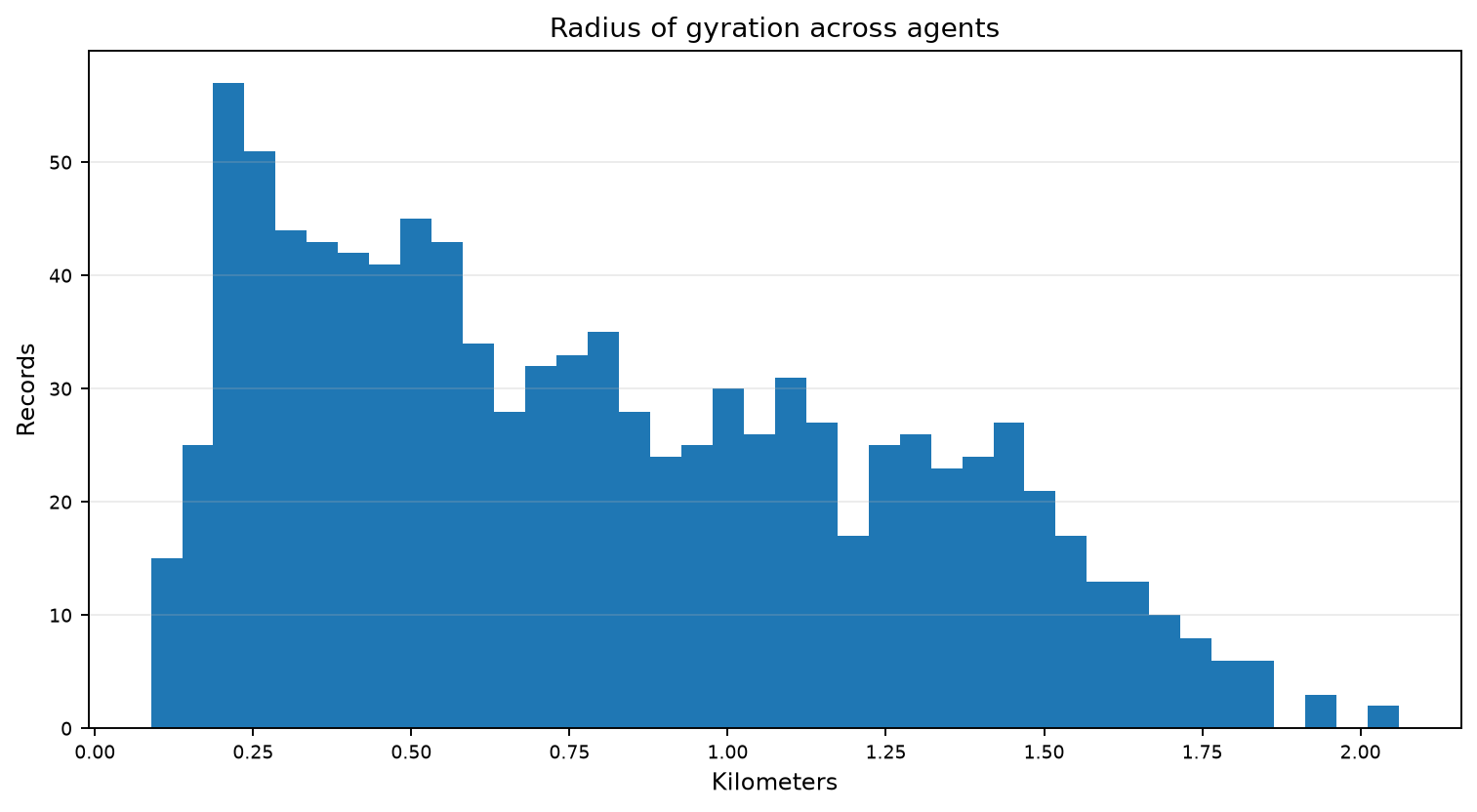}
\caption{Distribution of the radius of gyration across simulated agents.}
\label{fig:radius-of-gyration}
\end{figure}

\reffig{fig:radius-of-gyration} shows the distribution of agents' radius of gyration, which measures the typical spatial extent of an agent's movement around its center of activity. Most agents have a radius of gyration between approximately (0.2) and (0.6)~km, indicating that their daily activities are concentrated within relatively small geographic areas. The distribution is right-skewed, with progressively fewer agents exhibiting larger movement ranges. A smaller number of agents have radii between (1.0) and (1.6)~km, while only a few exceed (1.8)~km. This pattern suggests that the simulation produces a population dominated by localized mobility while still representing agents with broader travel ranges. We note that this distribution of radii of gyration closely resembles the distribution that has been reported in the literature for real-world taxi mobility data~\cite{wang2017exploring}. Compared to this study, we observe a similar peak at about low distances followed by a similar exponential decay to the right. A difference is that our distribution ends at about 2km where the distribution in ~\cite{wang2017exploring} ends at about 60km which is due to the smaller region used in our simulation compared to the region of Tianjin, China used in~\cite{wang2017exploring}. We note that the peak reported in~\cite{wang2017exploring} is much larger (about 5km) compared to ours (about 0.25km to 0.5km) which is explained by our dataset including walking and biking distances, whereas the work in~\cite{wang2017exploring} uses taxi data, only.

\begin{figure}[t]
\centering
\includegraphics[width=0.8\linewidth,trim=0cm 0cm 0cm 0.7cm, clip]{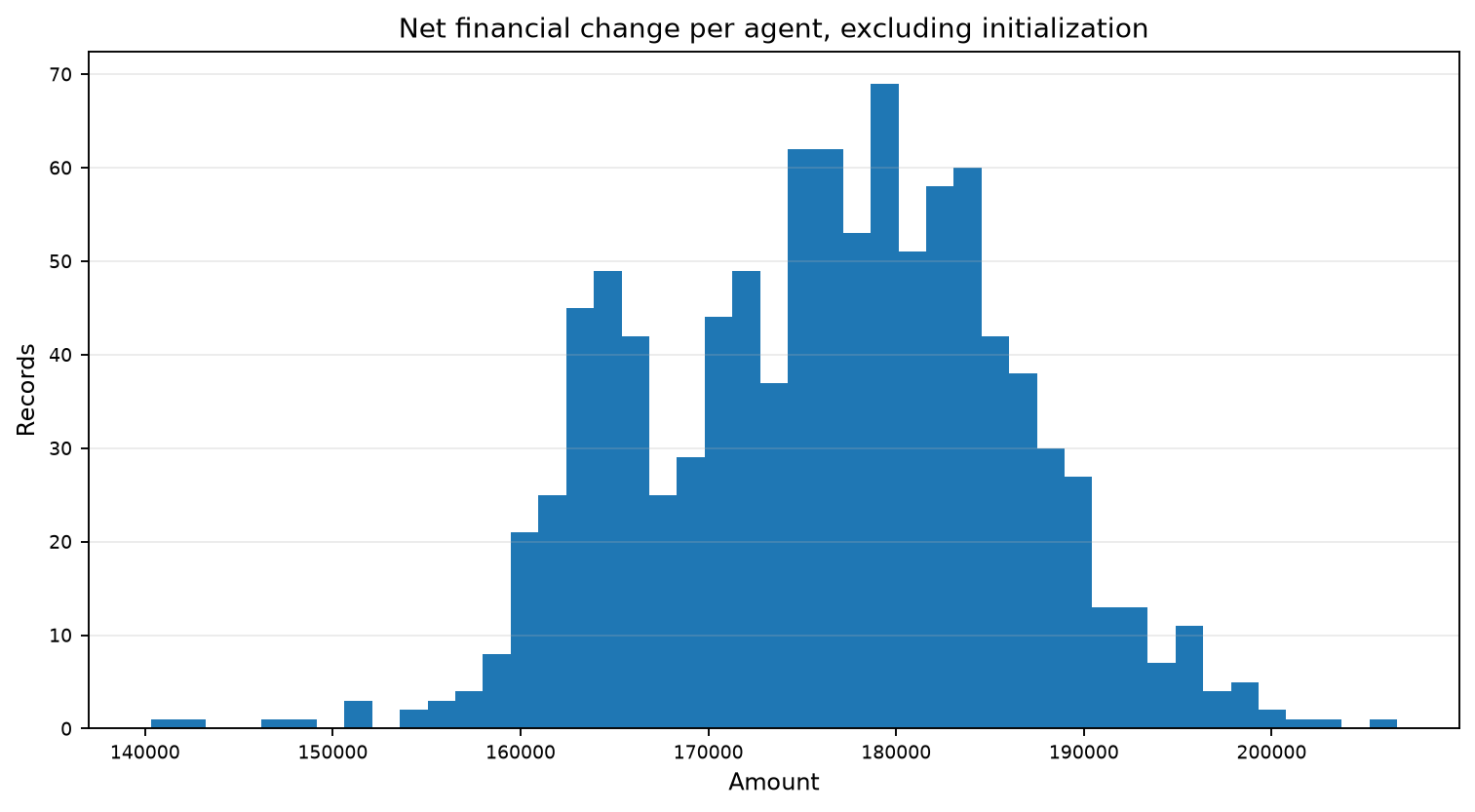}
\caption{Distribution of the net financial change per agent after excluding initialization-related transactions (simulation currency).}
\label{fig:agent-net-financial-change}
\end{figure}

\reffig{fig:agent-net-financial-change} shows the distribution of net financial change across agents after excluding transactions generated during initialization. Removing these initial transactions prevents starting balances and other setup-related transfers from distorting the analysis of the simulated economy. Most agents experience a net financial change between approximately (160{,}000) and (190{,}000) monetary units, with the highest concentration occurring near (180{,}000). The distribution is approximately unimodal, although moderate variation exists among agents because of differences in employment, income, expenditure, and activity patterns. Only a small number of agents appear at the lower and upper extremes. Overall, the concentration of values around a common range indicates that the simulation produces relatively stable financial outcomes while preserving agent-level economic heterogeneity.

% \begin{figure}[htbp]
% \centering
% \includegraphics[width=\linewidth]{figs/04_departure_time_by_destination.png}
% \caption{Hourly distribution of simulated trip start times by destination category. Values represent the percentage of trips within each destination category that begin during a given hour.}
% \label{fig:departure-time-by-destination}
% \end{figure}

% \reffig{fig:departure-time-by-destination} shows that trip timing varies substantially by destination. Work trips are strongly concentrated during the morning commuting period, reaching their highest share at approximately 7:00 and remaining elevated at 8:00 with a sharp drop after 9:00. A smaller secondary concentration appears around 12:00--15:00 which results from agents returning from lunch at restaurants. Trips with Home as the destination peak at approximately 16:00, reflecting the afternoon return-home period, and then gradually decline throughout the evening. Restaurant trips exhibit a midday peak between approximately 12:00 and 14:00, followed by a smaller evening increase around 17:00. Recreation trips are more broadly distributed across the afternoon and evening, with their highest shares occurring between approximately 16:00 and 18:00. Overall, these temporal patterns indicate that the simulation generates plausible emerging activity-specific schedules, including morning work departures, lunchtime restaurant visits, afternoon returns home, and evening recreational activity.

\begin{figure}[htbp]
    \centering

    \begin{minipage}[t]{0.49\textwidth}
        \centering
        \includegraphics[width=\linewidth]{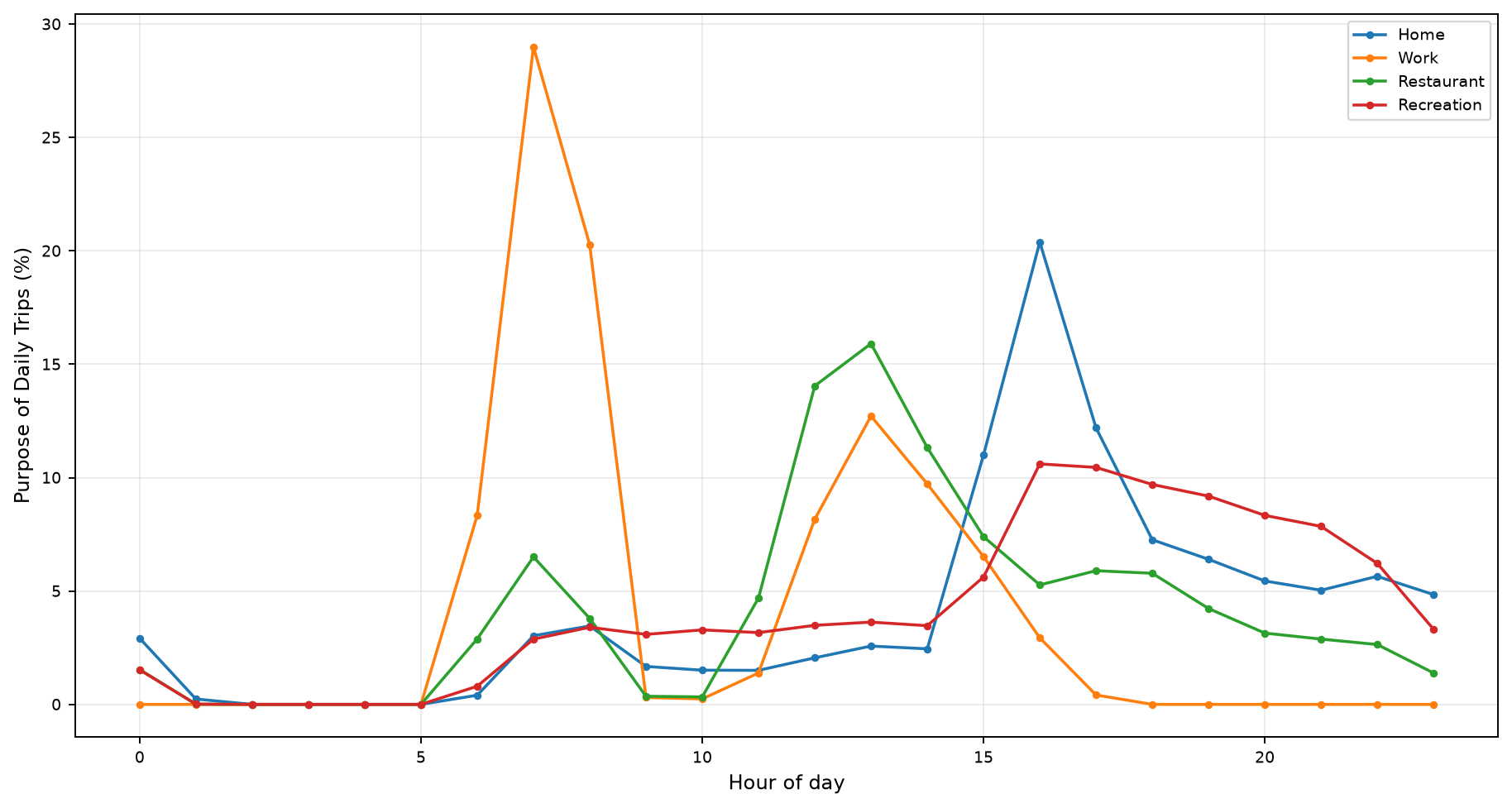}
        \par\smallskip
        \textbf{(a)} Trips by purpose in the simulation.
    \end{minipage}
    \hfill
    \begin{minipage}[t]{0.49\textwidth}
        \centering
        \includegraphics[width=\linewidth]{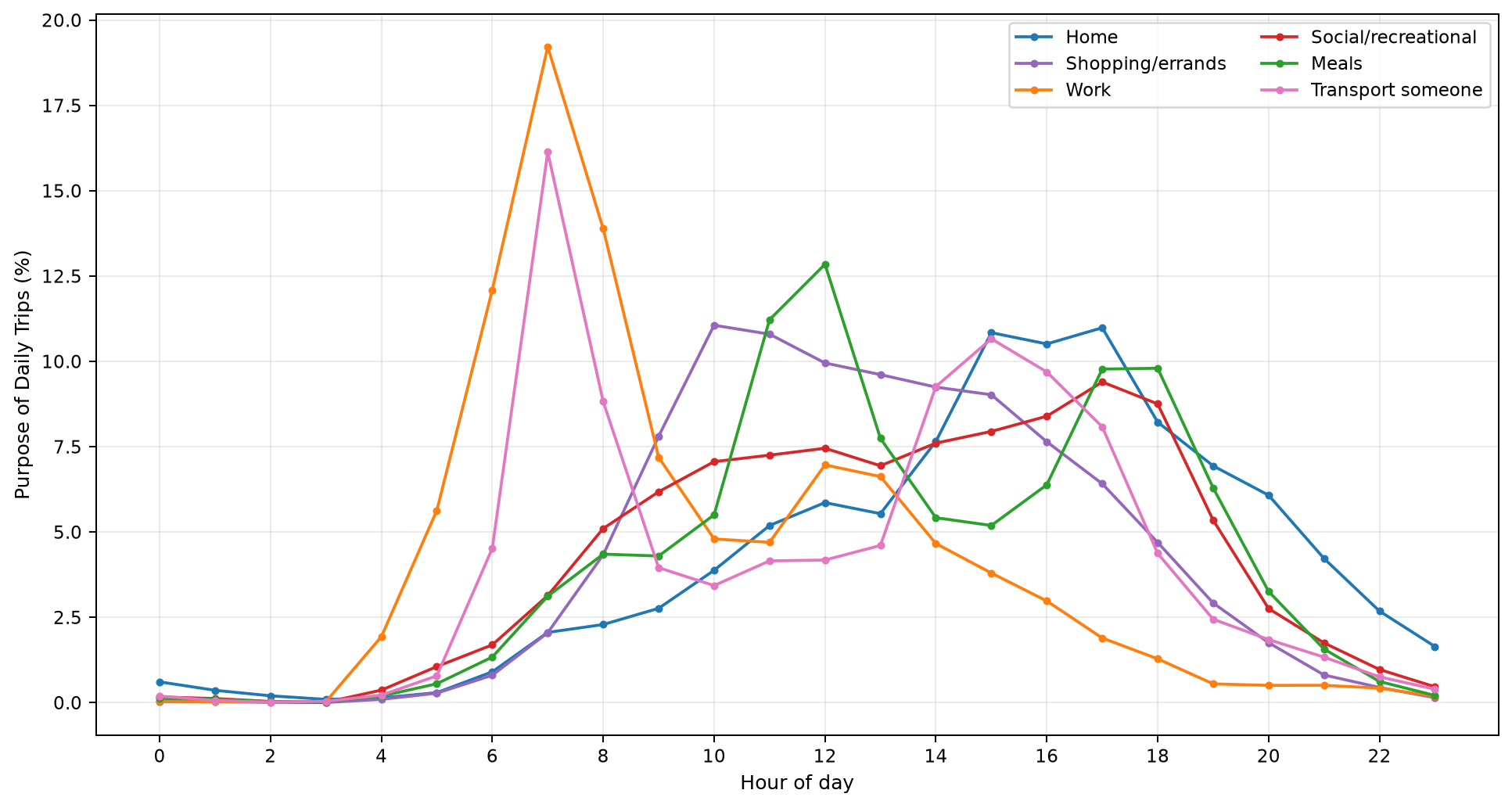}
        \par\smallskip
        \textbf{(b)} Trips by purpose in the 2017 NHTS.
    \end{minipage}

    \caption{Comparison of hourly trip departure patterns. (a) shows
    simulated trip start times by destination category, while (b)
    shows observed trip departure times by trip purpose in the 2017 National
    Household Travel Survey (NHTS). Values in each figure represent the
    percentage of daily trips within each category that begin during a given
    hour.}
    \label{fig:departure-time-comparison}
\end{figure}

\reffig{fig:departure-time-comparison} compares the temporal distribution of
simulated trips with the corresponding activity patterns observed in the 2017
National Household Travel Survey (NHTS). In the simulation
(\reffig{fig:departure-time-comparison}(a)), work trips are strongly
concentrated during the morning commuting period, reaching their highest share
at approximately 7:00 and remaining elevated at 8:00 before declining sharply
after 9:00. A smaller concentration around 12:00--15:00 is associated with
agents returning to work after lunch. Trips with Home as the destination peak
during the late afternoon, while restaurant trips exhibit a midday peak and a
smaller increase around the evening meal period. Recreation trips are more
broadly distributed across the afternoon and evening.
The NHTS distributions in \reffig{fig:departure-time-comparison}(b) show
similar activity-specific patterns. Work trips have a pronounced peak at
approximately 7:00, while trips made to transport another person also exhibit
a strong morning peak and a smaller afternoon concentration. Shopping and
errand trips are concentrated primarily between approximately 9:00 and 15:00,
with their highest shares occurring during the late morning. Meal-related
trips peak around noon and show a second increase between approximately 17:00
and 18:00. Social and recreational trips increase gradually throughout the
day and reach their highest shares during the late afternoon, whereas
home-bound trips are most common between approximately 15:00 and 18:00.
Overall, the similarity between the simulated and observed distributions
supports the plausibility of the emergent schedules, particularly the morning
work peak, midday meal activity, late-afternoon return-home period, and evening
recreational activity.

\section{Generated Dataset}
\label{sec:generated_dataset}
In this section, we present example datasets that were generated and processed using our software and have been previously published. Datasets smaller than 100 GB are hosted on OSF.io, with access links provided through our GitHub repository. For datasets exceeding 100 GB, we instead provide detailed instructions in the GitHub documentation to guide researchers in reproducing the data locally. This approach ensures that even very large datasets remain accessible while also allowing users to scale the simulation further by increasing the number of agents or extending the simulation duration.
Most datasets described in this work were generated using a compact desktop machine equipped with a 2.40 GHz Intel i5-1135G7 processor (8 cores, 16 GB RAM) running Linux Fedora. This demonstrates that the framework is efficient and does not require high-performance computing resources to produce large-scale mobility datasets, making it broadly accessible to both academic and industry researchers.

\subsection{Geolife+: Calibrated Dataset}

To demonstrate the scalability of our framework, we generated multiple datasets based on the GeoLife trajectories, varying both the number of simulated agents and the simulation duration. These datasets range from small-scale runs with hundreds of agents to large-scale scenarios involving tens of thousands of agents, spanning months to several years. Compared to the original GeoLife dataset, our simulations capture a richer set of agent activities, as every agent is continuously observed throughout the simulation period. This produces datasets that are significantly larger in size while maintaining realistic mobility characteristics, offering researchers an expanded view of population-scale mobility~\cite{amiri2024geolife+}.

\subsection{Massive Trajectory Data}
We generated a collection of large-scale check-in, location-based social network (LBSN) and trajectory datasets across four diverse urban and suburban regions: Fairfax (George Mason University campus area), New Orleans (French Quarter), San Francisco, and Atlanta. Each dataset integrates three components: simulated check-ins, evolving social links, and trajectory points recorded at five-minute intervals for every agent. Simulations were conducted at varying scales, from 1,000 to 10,000 agents over 15-month periods, as well as long-term runs spanning up to 20 years with 1,000 agents. These datasets capture both short-term dynamics and long-term behavioral patterns, providing a flexible foundation for mobility, social network, and anomaly detection research~\cite{amiri2023massive}.

\subsection{Human Mobility Dataset with Anomalies}

\begin{figure}
\centering
\includegraphics[trim = 0.0cm 0.0cm 0.0cm 0.0cm , clip,width=0.80\linewidth, height=8cm]{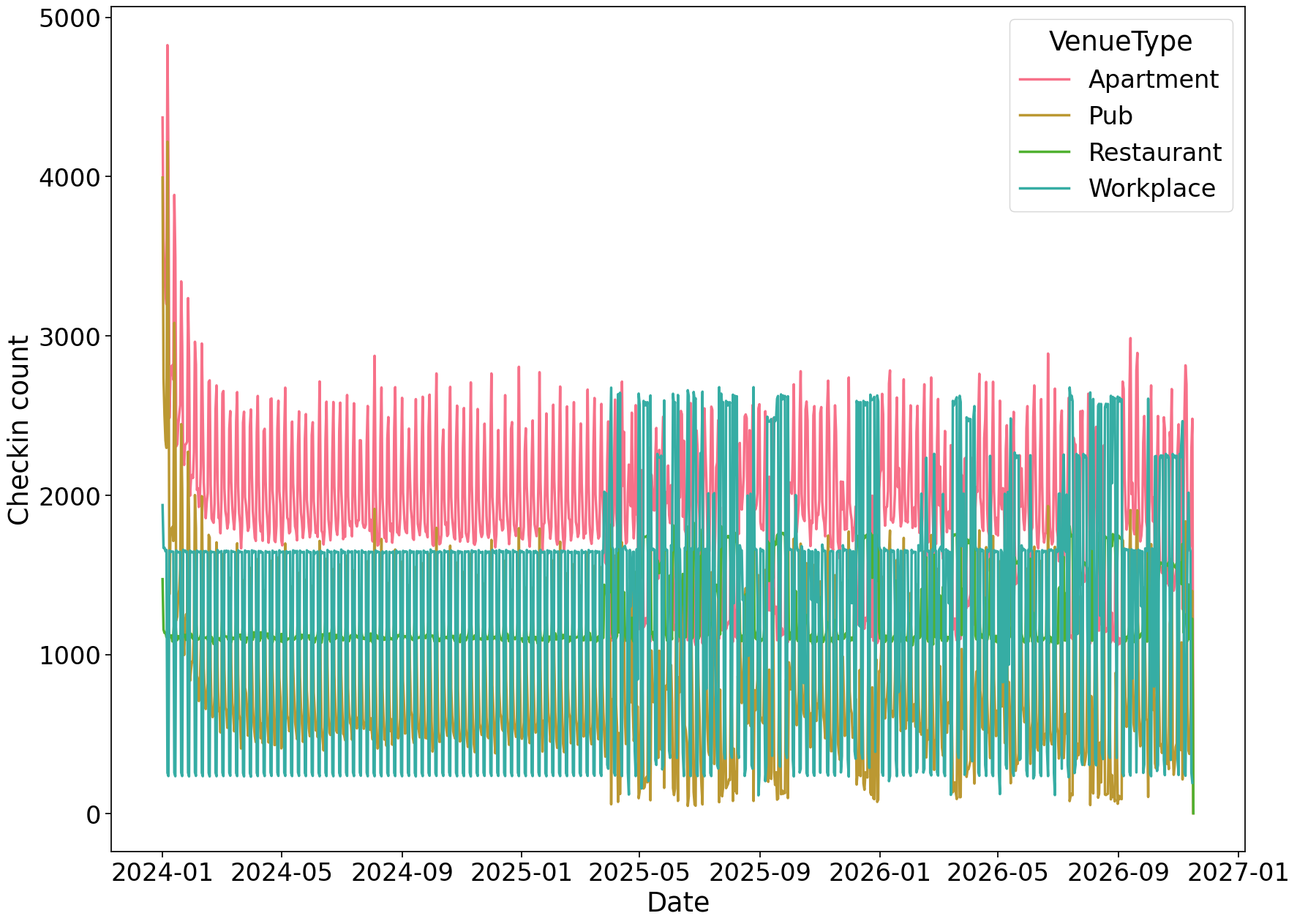}
\caption{Example of visualization phase output for simulated check-in data with global anomalies introduced over a 1050-day period}
\label{fig:anomaly-example}
\end{figure}

To demonstrate the flexibility of HD-GEN, we generated datasets that incorporate controlled anomalies designed specifically for anomaly detection research. The simulation environment allows users to inject a wide range of behavioral deviations into agent activities, including irregular eating patterns, atypical social interactions, and disruptions to routine work behavior. Each anomaly can be configured at different intensity levels and applied either to individual agents or to the entire population, enabling the creation of diverse and realistic experimental scenarios.
Agents can be designated as anomalous through three complementary mechanisms: (i) random assignment, in which anomalies are stochastically applied to selected agents; (ii) infectious disease style propagation, where anomalous behavior spreads dynamically through agent interactions, mimicking epidemiological processes; and (iii) location based triggers, where agents begin exhibiting anomalies after visiting specific locations. These mechanisms support the study of both isolated and systemic anomaly patterns and enable investigation into how abnormal behaviors emerge, propagate, and are detected in complex mobility systems~\cite{amiri2024urban}.

In related work, we introduced specific anomaly types to generate datasets for global anomaly detection tasks~\cite{zhang2024large,zhang2024transferable,liu2024neural}. The implemented anomaly categories include hunger anomalies, social anomalies, and work anomalies. Hunger anomalies cause agents to experience hunger more frequently than normal, resulting in increased check ins at restaurants or homes. Social anomalies cause agents to ignore usual preferences and social influences, selecting recreational locations at random. Work anomalies disrupt routine behavior by causing agents to skip work on days when attendance is expected.
Each anomaly type is further parameterized by three intensity levels: red, orange, and yellow. Red level anomalies represent extreme behavioral deviations, such as skipping work 100\% of the time, orange level anomalies represent moderate deviation at approximately 50\%, and yellow level anomalies represent mild deviation at approximately 20\%. Anomalies can be applied either to specific agents for a defined duration or globally across all agents on selected days.

Using the visualization tools described in \ref{sec:visualization}, \reffig{fig:anomaly-example} presents an example of a global anomaly experiment. The figure shows check in patterns for multiple venue types, including apartments, pubs, restaurants, and workplaces, over a 1050 day simulation period starting in January 2024. The first 450 days simulate normal behavior, allowing agents to establish stable baseline routines. During the remaining 600 days, global anomalies are introduced, where all agents adopt a specific anomaly type and intensity level on each day.
As illustrated in the figure, agent behavior remains stable during the baseline phase and becomes increasingly irregular once anomalies are introduced. For example, during red level global hunger anomaly events, all agents experience heightened hunger, leading to increased visits to restaurants and homes. This produces sharp spikes in restaurant and home check ins, accompanied by fluctuations in workplace visits as agents leave work to eat.

We used the provided tools to generate the dataset reported in~\cite{amiri2024urban}, which contains multiple anomaly types and configurations. The framework further supports flexible control over anomaly distribution by allowing users to select between random assignment, infectious propagation, and location based spreading mechanisms. Together, these capabilities enable the generation of rich, realistic datasets for evaluating anomaly detection methods across a wide range of behavioral and mobility scenarios.

% \section{Dataset Re-Generation Instructions}
% \label{sec:reproduction}
% \input{content/reproduction.tex}

\section{Conclusions}
\label{sec:conclusions}
In this paper, we have presented a software system to generate, calibrate and process large-scale synthetic geospatial datasets by simulating the pattern of life of individuals with the consideration of real-world constraints. We have shown the effectiveness of our approach by generating a set of synthetic datasets based on the GeoLife dataset. Furthermore, we have shown that the generated datasets exhibit similar statistical properties to the original dataset, and can be used for various geospatial data analysis tasks. We have also provided detailed instructions on how to reproduce the generated datasets, and have made the code and data available on GitHub. With the provided datasets and code, researchers can easily generate large-scale synthetic geospatial datasets for their research purposes and evaluate the performance of their algorithms on realistic data.
 Our approach bridges the gap between real-world and simulated trajectory data by integrating statistical features from real datasets with the scalability of simulations. By introducing a data generation tool, a genetic algorithm for calibration, and a data processing pipeline, we enable the creation of realistic, large-scale mobility datasets suitable for diverse applications. 
 This framework provides a robust foundation for research in human mobility modeling, machine learning, and data-driven applications while addressing the limitations of both real and synthetic datasets. %\newpage

 Future work will focus on improving scalability to support simulations involving millions of individuals, incorporating transportation-related constraints, and supporting modern geospatial data formats such as GeoParquet to improve storage efficiency, interoperability, and scalability. The current use of shapefiles is primarily maintained for compatibility with POL, which presently accepts shapefiles as input.
\section{Acknowledgments}
\subsection*{Funding}
Supported by the Intelligence Advanced Research Projects Activity (IARPA) via Department of Interior/ Interior Business Center (DOI/IBC) contract number 140D0423C0025. The U.S. Government is authorized to reproduce and distribute reprints for Governmental purposes notwithstanding any copyright annotation thereon. Disclaimer: The views and conclusions contained herein are those of the authors and should not be interpreted as necessarily representing the official policies or endorsements, either expressed or implied, of IARPA, DOI/IBC, or the U.S. Government.

This research was also supported by the National Science Foundation under Award Abstract \#2109647, Data-Driven Modeling to Improve Understanding of Human Behavior, Mobility, and Disease Spread. 

\subsection*{Use of Language Assistance Tools}
We clarify that to write this paper we used ChatGPT 5.x only to improve the English and readability of our writing. We did not ask it to write any sections or paragraphs from its own knowledge.
\bibliographystyle{ACM-Reference-Format}
\bibliography{main}

\end{document}